\documentclass[american,authoryear, 12]{elsarticle}
\usepackage[T1]{fontenc}
\usepackage[latin9]{inputenc}
\usepackage{xcolor}
\usepackage{pdfcolmk}
\usepackage{array}
\usepackage{float}
\usepackage{units}
\usepackage{textcomp}
\usepackage{mathrsfs}
\usepackage{amsmath}
\usepackage{graphicx}
\usepackage{setspace}
\PassOptionsToPackage{normalem}{ulem}
\usepackage{ulem}

\makeatletter

\providecommand{\tabularnewline}{\\}

\floatstyle{ruled}
\newfloat{algorithm}{tbp}{loa}
\providecommand{\algorithmname}{Algorithm}
\floatname{algorithm}{\protect\algorithmname}
\providecolor{lyxadded}{rgb}{0,0,1}
\providecolor{lyxdeleted}{rgb}{1,0,0}

\usepackage{hyperref}

\makeatother

\usepackage{babel}
\begin{document}

\begin{frontmatter}{}

\title{Efficiency of BRDF sampling and bias on the average photometric behavior}

\author{Fr\'ed\'eric Schmidt $^{1}$, S\'ebastien Bourguignon$^{2}$}

\address{$^{1}$ GEOPS, Univ. Paris-Sud, CNRS, Universit\'e Paris-Saclay,
Rue du Belv\'ed\`ere, B\^at. 504-509, 91405 Orsay, France, $^{2}$
LS2N, \'Ecole Centrale de Nantes, CNRS, 1 rue de la No\"e, 44321
Nantes, France}
\begin{abstract}
The Hapke model has been widely used to describe the photometrical
behavior of planetary surface through the Bi-directional Reflectance
Distribution Function (BRDF), but the uncertainties about retrieved
parameters has been difficult to handle so far. A recent study proposed
to estimate the uncertainties using a Bayesian approach (Schmidt et
al., Icarus 2015). In the present article, we first propose an improved
numerical implementation to speed up the uncertainties estimation.
Then, we conduct two synthetic studies about photometric measurements
in order to analyze the influence of observation geometry:

First, we introduce the concept of ``efficiency'' of a set of geometries
to sample the photometric behavior. A set of angular sampling elements
(noted as geometry) is efficient if the retrieved Hapke parameters
are close to the expected ones. We compared different geometries and
found that the principal plane with high incidence is the most efficient
geometry among the tested ones. In particular, such geometries are
better than poorly sampled full BRDF.

Second, we test the analysis scheme of a collection of photometric
data acquired from various locations in order to answer the question:
are these locations photometrically homogeneous or not? For instance,
this question arises when combining data from an entire planetary
body, where each spatial position is sampled at a single geometry.
We tested the ability of the Bayesian method to decipher two situations,
in the presence of noise: (i) a photometrically homogeneous surface
(all observations with the same photometric behavior), or (ii) an
heterogeneous surface with two distinct photometrical properties (half
observations with photometric behavior 1, other half with photometric
behavior 2). We show that the naive interpretation of the results
provided by Bayesian method is not able to solve this problem. Therefore,
we propose a separability test based on chi-square analysis, which
is able to distinguish the two situations and thus, to access information
about the photometric heterogeneity of a planetary body. When the
noise level is high (10 \%), our simulations show that separability
cannot be solved.
\end{abstract}

\end{frontmatter}{}


\section{Introduction}

Photometry is the study of the light reflected by the surface/atmosphere
of a planetary body. The Bi-directional Reflectance Distribution Function
(BRDF) is the core quantity to describe the photometric behavior \citep{Hapke_Book1993}.
An Emission Phase Function (EPF) is a particular case of BRDF when
the incidence direction is fixed, and the emergence directions are
sampled along one plane. Hereafter, we note ``geometry'' the set
of incidence and emergence directions.

Hapke proposed a model of the BRDF for a granular medium \citep[e.g.][]{Hapke_BRDF1theroy_JGR1981,Hapke_BRDF2Experiments_JGR1981,Hapke_BRDF3roughness_Icarus1984,Hapke_BRDF4ExtinctionOpposition_Icarus1986,Hapke_BRDF5scatterringCBOE_Icarus2002,Hapke_BRDF6_Porosity_Icarus2008}.
Even if this model has been controversial \citep[e.g.][]{Mishchenko1994,Hapke_PlanetaryBackscatering_JQSRT1996,Shepard_testHapkephotometric_JoGRP2007,Shkuratov_criticalassessmentHapke_JoQSaRT2012,Hapke_CommentAcritical_JoQSaRT2013},
many authors have been using it to analyze laboratory data \citep[e.g.][]{Cord_PlanetBDRF_Icarus2003,Souchon_experimentalstudyof_I2011,Beck_Photometryofmeteorites_I2012,Pommerol_PhotometricpropertiesMars_JGRP2013,Johnson_Spectrogoniometryandmodeling_I2013,Pilorget_Wavelengthdependencescattering_I2016},
telescopic observations \citep[e.g.][]{hapke_oppostition_Icarus1998},
in situ data \citep[e.g.][]{Johnson_Preliminaryresultsphotometric_JGR1999,Johnson_Spirit_JGR2006,Johnson_Opportunity_JGR2006},
remote sensing data \citep[e.g.][]{Jehl_GusevPhotometry_Icarus2008,Yokota_Lunarphotometricproperties_I2011,Fernando_SurfacereflectanceMars_JoGRP2013,Vincendon_Marssurfacephase_PaSS2013,Sato_ResolvedHapkeparameter_JoGRP2014,Fernando_Characterizationandmapping_I2015,Fernando_MartianSurfaceMicrotexture_2016},
due to its relative simplicity and fast computation.

Recently, a new approach based on a Bayesian framework and a Monte-Carlo
Markov Chain (MCMC) algorithm has been proposed to estimate realistic
uncertainties on the estimated Hapke parameters from actual measurements
\citep{Schmidt_RealisticuncertaintiesHapke_I2015}. Following this
study, we do not discuss here the realism of the photometric Hapke
model, but focus on the data analysis point of view. The present article
aims at:
\begin{enumerate}
\item improving the numerical implementation of the MCMC algorithm \citep{Schmidt_RealisticuncertaintiesHapke_I2015}
to efficiently analyze photometric datasets but also spectro-photometric
datasets;
\item exploiting the outputs of this algorithm in order to determine the
optimized observation geometries in BRDF and EPF geometries;
\item exploiting the algorithm outputs to try to distinguish between homogeneous
and heterogeneous photometric datasets for any geometry.
\end{enumerate}
This work should help to interpret previous analyses but also to design
future instrumental and observational campaigns \citep{Nag_Observingsystemsimulations_IJoAEOaG2015}.
The second point has been firstly addressed using a standard least-squares
estimation method \citep{1989aste.conf..557H}. Thanks to a more sophisticated
data analysis, the Bayesian method provides more information than
the latter, allowing in particular robust error estimation. 

In natural scene observation, a mixture of materials, unresolved at
the pixel scale, could lead to non-linear effects \citep{Pilorget_Photometryparticulatemixtures_I2015},
sometimes difficult to handle. The third point aims at discussing
another aspect of the difficulty: the heterogeneity of the dataset.
For example, this question arises when analyzing \emph{in situ} rover
data: one could assemble measurements records from various sites at
different geometries, assuming that they have the same photometric
behavior. Also, one could assemble the data from disk-resolved images,
in order to estimate the global photometric behavior of a body. In
both cases, photometric heterogeneity may affect the dataset. We propose
and quantify here an approach to decide if the dataset can be considered
as homogeneous or not.

\section{Method}

In this section, we first briefly describe the Hapke photometric model.
Then, an improvement of the MCMC sampling algorithm is proposed. The
next part introduces an indicator of the efficiency of a set of geometries
to retrieve the correct parameters. Finally, the methods to decipher
the homogeneity/heterogeneity of the dataset are presented.

\subsection{Hapke's photometric model \label{subsec:Direct-model}}

We use the standard 1993 Hapke model of bidirectional reflectance
\citep{Hapke_Book1993}, that has been commonly used in the planetary
science community \citep{Johnson_Spirit_JGR2006,Johnson_Opportunity_JGR2006,Jehl_GusevPhotometry_Icarus2008,Beck_Photometryofmeteorites_I2012,Fernando_SurfacereflectanceMars_JoGRP2013,Fernando_Characterizationandmapping_I2015,Pilorget_Wavelengthdependencescattering_I2016},
described by the following equation:

\begin{equation}
r=\frac{\omega}{4\pi}\,\frac{\mu_{0e}}{\left(\mu_{0e}+\mu_{e}\right)}\,\left\{ \left[1+B(g)\right]P(g)+H(\mu_{0e})H(\mu_{e})-1\right\} S(\theta_{0},\bar{\theta},g).\label{eq:Hapke}
\end{equation}
\\
We express the reflectance $r$ in REFF (REFlectance Factor) unit:

\begin{equation}
\textrm{REFF}=r\frac{\pi}{\mu_{0}},\label{eq:REFF}
\end{equation}
\\
that can be normalized by the main geometrical effect:

\begin{equation}
\textrm{REF\ensuremath{\mathrm{F}_{norm}}}=\textrm{REFF}(\mu_{0}+\mu).\label{eq:REFF-normalized}
\end{equation}

We use the same notations and definitions as in \citealt{Schmidt_RealisticuncertaintiesHapke_I2015}.
The main parameters are recalled here:
\begin{itemize}
\item $\theta_{0}$, $\theta$, and $g$: incidence, emergence and phase
angles, respectively (subscript $_{e}$ denotes the equivalent geometry
in the rough case as defined in the Hapke model). The whole geometry
quantities are noted $\Omega=(\theta_{0},\theta,g)$. $\varphi$ is
the azimuth angle. $\mu=\cos\theta$ and $\mu_{0}=\cos\theta_{0}$
are defined to simplify the expressions.
\item $\omega$ ($0\le\omega\le1$): single scattering albedo. It represents
the fraction of scattered to incident radiation by a single particle
(sometimes noted w).
\item $P(g)$: particle scattering phase function described by the 2-terms
Henyey-Greenstein function \citep{Henyey_DiffusionGalaxy_ApJ1941}
(hereafter referred to as HG2):
\begin{equation}
P(g)=\left(1-c\right)\,\frac{1-b^{2}}{\left(1+2b\cos\left(g\right)+b^{2}\right)^{3/2}}+c\,\frac{1-b^{2}}{\left(1-2b\cos\left(g\right)+b^{2}\right)^{3/2}}\label{eq:HG2}
\end{equation}
The HG2 function depends on two parameters: the asymmetry parameter
$b$ ($0\le b\le1$) which characterizes the anisotropy of the scattering
lobe, and the backscattering fraction $c$ ($0\le c\le1$) which characterizes
the main direction of the diffusion.
\item $H(x)$: multiple scattering function. With $y=(1-\omega)^{1/2}$,
the multiple scattering function is \citep{Hapke_Book1993}:
\begin{equation}
H(x)=\left\{ 1-\left[1-y^{2}\right]x\left[\left(\frac{1-y}{1+y}\right)+\left(\frac{1}{2}-x\left(\frac{1-y}{1+y}\right)\right)\ln\left(\frac{1+x}{x}\right)\right]\right\} .^{-1}\label{eq:H1993}
\end{equation}
A new expression dedicated to anisotropic scattering has been proposed
in \citealp{Hapke_BRDF5scatterringCBOE_Icarus2002}. Nevertheless,
\citealt{Pommerol_PhotometricpropertiesMars_JGRP2013} noticed that
the use of the latter expression does not lead to any significant
change with respect to the former one.
\item $B(g)$: opposition effect function. The Shadow Hiding Opposition
Effect (SHOE) is taken into account as follows \citep{Hapke_Book1993}:
\begin{equation}
B(g)=\frac{B_{0}}{1+\frac{1}{h}\tan\left(\frac{g}{2}\right)}.
\end{equation}
This function depends on the parameters $h$ and $B_{0}$ (ranging
from 0 to 1), which respectively denote the angular width and the
amplitude of the opposition effect.
\item $S(\theta_{0},\bar{\theta},g)$: macroscopic roughness factor given
in \citealp{Hapke_Book1993}, depending on the roughness parameter
$\bar{\theta}$ and geometry ($\theta_{0},g$).
\end{itemize}
Thus, we can summarize the Hapke model as a function, depending on
geometries and surface properties:

\begin{equation}
r\left((\theta_{0},\theta,g),(\omega,b,c,\bar{\theta},h,B_{0})\right)
\end{equation}
Hereafter, for a set of geometries $\Omega_{i}=(\theta_{0},\theta,g)$
and a set of parameters $m=(\omega,b,c,\bar{\theta},h,B_{0})$, we
denote the synthetic reflectance by $r\left(\Omega_{i},m\right)$.

\subsection{Improved Monte-Carlo Markov Chain (MCMC) algorithm\label{subsec:MCMC}}

Because the Hapke model (eq. \ref{eq:Hapke}) is non-linear, the inverse
problem consisting in estimating parameters $m=(\omega,b,c,\bar{\theta},h,B_{0})$
from a set of measured reflectances $r_{i}$ at different geometries
is a difficult question. In particular, least-squares fitting by dedicated
optimization may converge toward a local, unsatisfactory, minimum
point. On the contrary, the probabilistic inversion setting, combined
with appropriate numerical methods, makes it possible to handle complex
solutions \citep{Tarantola_QuestInformation_JGeophys1982,Mosegaard_MonteCarloInversion_JGR1995}.
By describing all quantities in terms of Probability Density Functions
(PDFs), it also provides a solid framework to incorporate some prior
information on the unknown parameters (Bayesian model) and to derive
uncertainties on the estimated parameters. We recall here the most
important quantities involved in the computations. Notations are similar
to those of the previous article \citep{Schmidt_RealisticuncertaintiesHapke_I2015}.
The quantity $\sigma_{M}(m)$ represents the final solution in the
parameter space $M$, that is, the joint posterior PDF of all parameters:

\begin{equation}
\sigma_{M}(m)=k\,L(m)\,\rho_{M}(m),\label{eq:PDF}
\end{equation}
where, according to Bayes' rule, $L(m)$ is the likelihood function,
$\rho_{M}(m)$ is the prior distribution on the model parameters and
$k$ is a normalizing constant such that $\int\sigma_{M}(m)dm=1$. 

We consider a Gaussian error on each measured reflectance $r_{i}$,
with standard deviation. The likelihood of a collection of observations
$r_{i}$, sampled at geometries $\Omega_{i}$ , then reads:
\begin{equation}
L=\left(\prod_{i}\frac{1}{\sigma_{i}\sqrt{2\pi}}\right)\exp\left\{ -\frac{1}{2}\sum_{i}\frac{\left(r_{i}-r\left(\Omega_{i},m\right)\right)^{2}}{\sigma_{i}^{2}}\right\} \label{eq:Likelyhood}
\end{equation}

In this paper, uniform prior distributions are considered in the parameter
space, that is, in $\left[0,1\right]$ for all parameters, except
for $\bar{\theta}$ which is restricted to in $\left[0\text{\textdegree},45\text{\textdegree}\right]$.
Consequently, the posterior distribution is proportional to the likelihood
function within the definition domain of the parameters. Let us remark
that similar methodology could be developed in cases where more informative
prior distributions are available.

The PDF $\sigma_{M}(m)$ is sampled by a collection of $N_{samp}=10^{5}$
vectors, noted as:
\begin{equation}
\check{m_{\ell}},\:\ell=1,\ldots,N_{samp}.
\end{equation}
\\
Each sample vector $\check{m_{\ell}}$ has a corresponding reflectance
vector $\check{r_{\ell}},$ obtained by computing the Hapke model
with parameters $\check{m_{\ell}}$. Since the expression of the PDF
\eqref{eq:PDF} is complex, obtaining such samples requires a specifically
designed procedure, namely, a Monte-Carlo Markov Chain (MCMC) algorithm.

In the previous implementation of the MCMC method \citep{Schmidt_RealisticuncertaintiesHapke_I2015},
a naive distribution was used (samples were drawn uniformly in the
parameter space), which resulted in a very high rejection rate and
lacked efficiency, especially in the case of strongly constrained
solutions: if the PDF of interest $\sigma_{M}$ is spiky, that is,
sharply concentrated around its mode(s), then most samples are drawn
in low probability regions and are therefore rejected. In this paper,
more efficient distributions are proposed, which improve the algorithmic
efficiency for either strongly and weakly constrained cases. More
precisely, we use a mixture of proposal distributions: for each parameter,
a new sample is drawn either from a uniform distribution over the
parameter space (with probability 1/5), or from a Gaussian random
walk centered on the current parameter value (and restricted to the
definition domain) with high standard deviation (with probability
2/5), or with low standard deviation (with probability 2/5). Then,
the candidate is accepted with a probability given by the Metropolis-Hastings
ratio \citep{Robert_MonteCarlo_book2005}, otherwise the previous
value is repeated, ensuring that the distribution of the samples asymptotically
follows that in \eqref{eq:PDF}. Doing so improves the exploration
of the parameter space, since it combines large-scale moves (two first
cases) enabling jumps between possible local modes, and small-scale
moves (last case) for refined, local, exploration \citep{Andrieu_JointBayesianmodel_IToSP1999}.
Note that such a mix between different distributions does not impact
convergence properties of the sampling algorithm.

In practice, an initial set of parameters is drawn uniformly in the
parameter space. Consequently, the first iterations of the MCMC algorithms
are usually discarded since they depend on initialization (burn-in
period). In all our experiments, we generated 100,000 samples, from
which the first 5000 ones were discarded, which always appeared to
be sufficient.

The pseudo-code of the algorithm is given in \ref{sec:Appendix-Monte-Carlo-Markov}. 

\subsection{Measure of the efficiency of a set of geometries}

We propose here a numerical method to determine the efficiency of
a set of geometries to estimate the Hapke parameters, for an unknown
material following \citealp{1989aste.conf..557H}. We consider the
Hapke BRDF model, with four possible parameters: $\omega$, $b$,
$c$, $\bar{\theta}$. Since $B_{0}$ and $h$ are known to be constrained
with small phase angles geometries, we do not consider them in the
analysis. In order to quantify the ability of a set of geometries
to retrieve the correct Hapke parameters, we generate an artificial
dataset of reflectances obtained according to the Hapke model for
a set of ``true'' parameters $(\omega_{1},b_{1},c_{1},\bar{\theta}_{1})$,
with a realistic noise level. Then, the parameter distribution $\sigma_{M}$
is estimated by the algorithm described in Section \ref{subsec:MCMC}.
The closer the solution to the true parameter set, the better the
estimate. We measure the quality of the tested geometry by considering,
for every parameter (for example $\omega$), the part of the distribution
$\sigma(\omega)$ which lies inside the interval $[\omega_{1}-\epsilon,\omega_{1}+\epsilon]$,
where $\epsilon$ is set to 1\% of the total parameter space, so a
margin of 2\% of the total parameter space is allowed (between 1\%
smaller and 1\% greater than the true value $\omega_{1}$). For $\bar{\theta}$,
$\epsilon$ is set to 0.45 since it represents angles from 0 to 45\textdegree .
Let us denote $I_{\omega}=\int_{\omega_{1}-\epsilon}^{\omega_{1+\epsilon}}\sigma(\omega)d\omega.$
In practice, $I_{w}$ is simply estimated by considering the proportion
of samples $\omega_{\ell}$ that fall inside the correct interval
among the $N_{samp}$ samples that were drawn. We then consider:

\begin{equation}
D_{\omega}=-\log I_{\omega}\label{eq:Kullback_Liebler_dirac-1}
\end{equation}
\\
so that $D_{\omega}$ decreases with $I_{\omega}$, equals 0 when
the full distribution $\sigma(\omega)$ lies inside the interval of
interest $[\omega_{1}-\epsilon,\omega_{1}+\epsilon]$ and equals $+\infty$
if $\sigma(\omega)=0$ in $[\omega_{1}-\epsilon,\omega_{1}+\epsilon]$.
Quantities $D_{b}$, $D_{c}$, $D_{\bar{\theta}}$ are defined similarly.

The total efficiency distance $E$ of a given geometry is finally
defined by the sum:

\begin{equation}
E=D_{\omega}+D_{b}+D_{c}+D_{\bar{\theta}}.\label{eq:Efficiency_geometry-1}
\end{equation}

\begin{figure}
\includegraphics[bb=0bp 0bp 561bp 510bp,clip,width=0.7\textwidth]{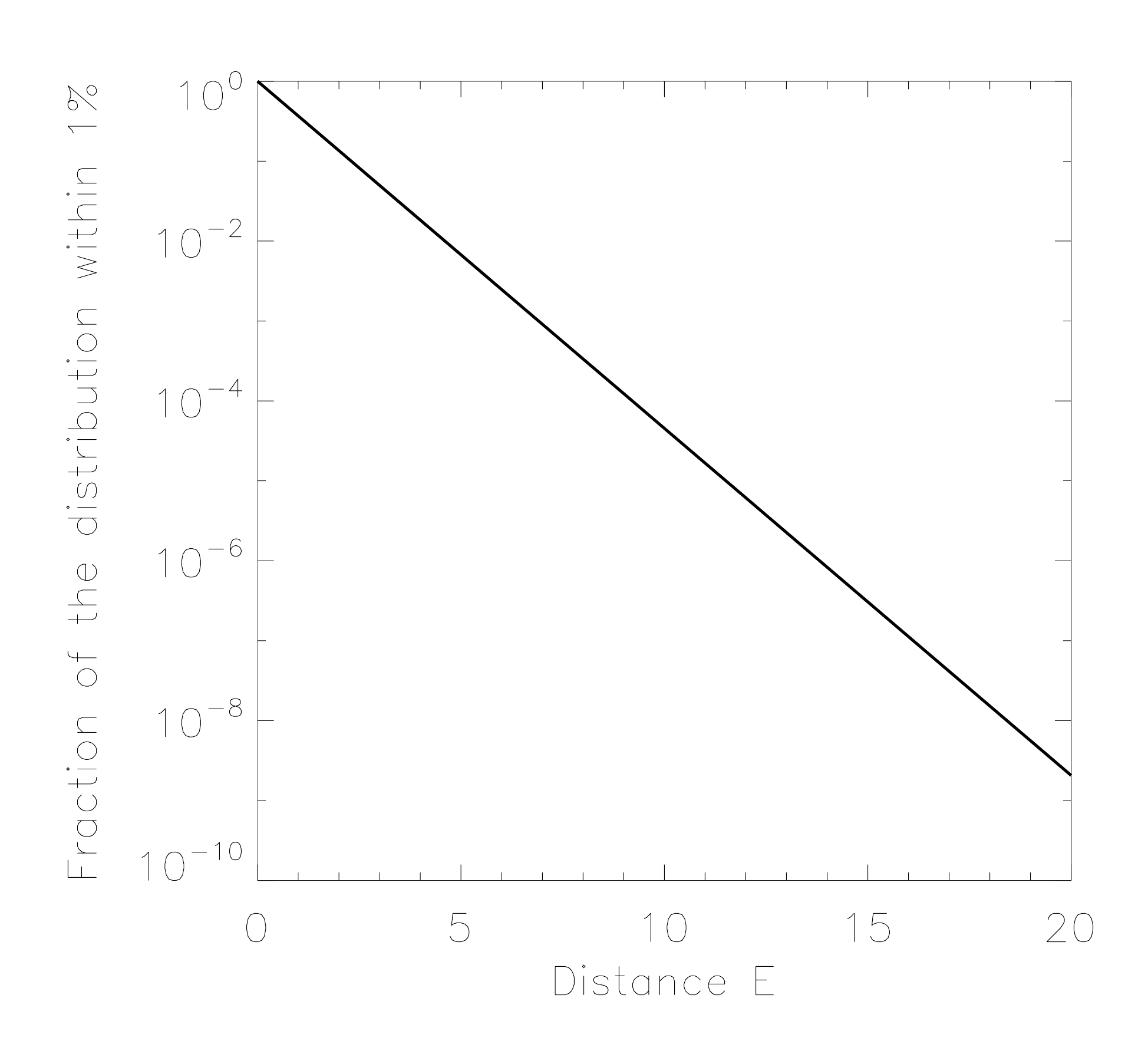}\centering

\caption{Accuracy of the retrieval as a function of efficiency distance $E$,
noted as a distance from Eq. \ref{eq:Efficiency_geometry-1}. If the
distance $E$ is zero, then the full estimated distribution lies in
an interval centered at the expected solution, with 1\% margin (between
1\% greater and 1\% smaller). If the distance $E=1$, then only 37\%
of the distribution is close to the expected solution within 1\% margin.
\label{fig:DistanceDistrib}}
\end{figure}

Figure \ref{fig:DistanceDistrib} shows the correspondence between
the efficiency $E$ and the fraction of the distribution in the close
neighborhood of the expected solution (2\% margin). When a set of
geometries is perfectly efficient to describe the photometric behavior,
the whole distribution is within the 2\% margin and the distance $E=0$.
When the geometry does not contain any information about the parameters,
the estimated PDF is uniform. Since we have 4 parameters, in this
case, we expect to have only a fraction of $0.02^{4}$ of the MCMC
samples in the 2\% margin. The corresponding distance is $E=15.6$.
Note that an even larger value may be obtained if less than 2\% of
the samples fall in the relevant domain, that is to say, when the
estimated distribution is even more unfavorable than the uniform,
non-informative, one.

\subsection{Strategy for separability\label{subsec:Strategy-for-separability}}

\begin{figure}
\hfill{}\includegraphics[width=0.4\textwidth]{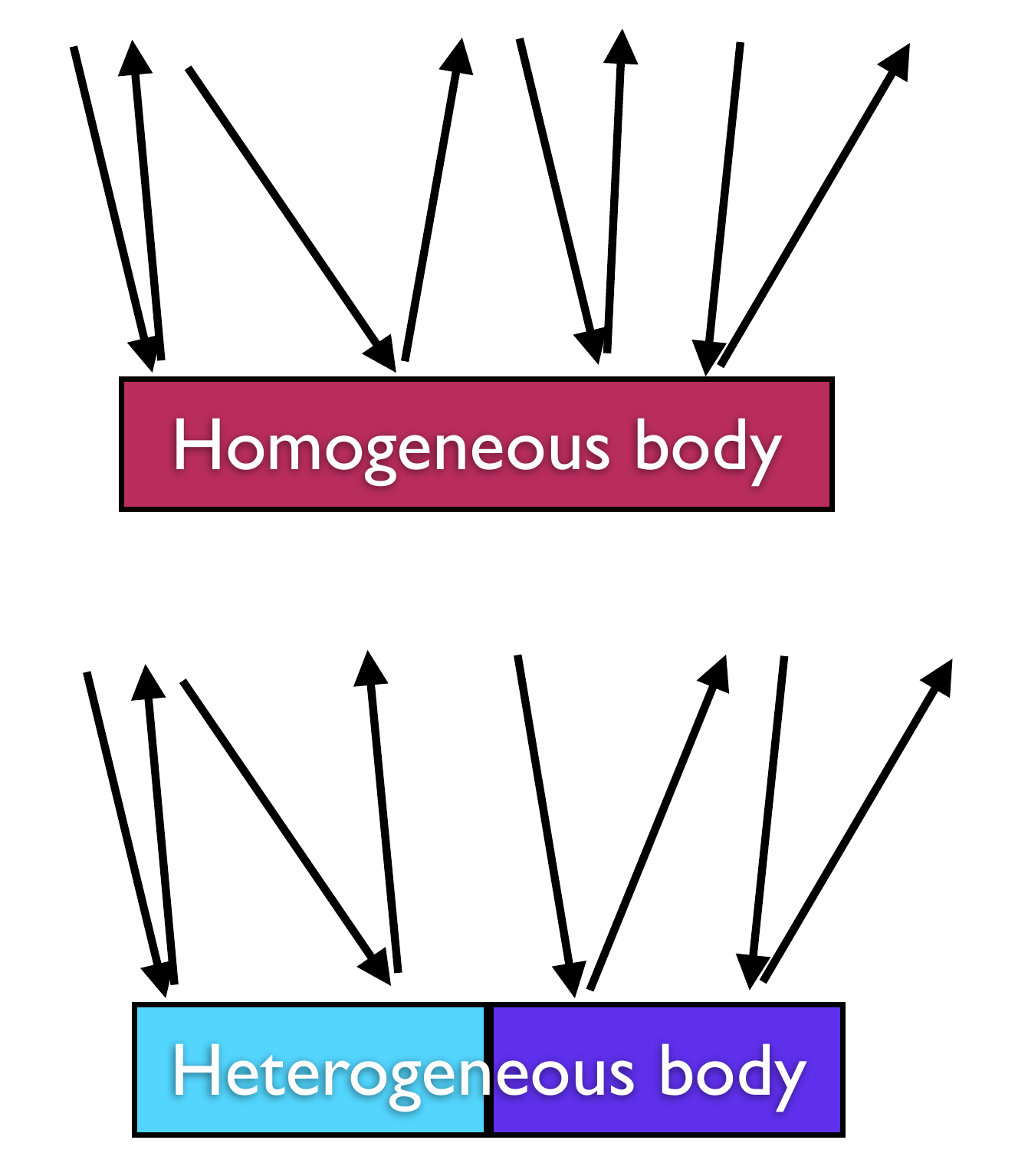}\hfill{}

\caption{Sketch of an observation of an homogeneous surface (top) and of an
heterogeneous surface (bottom). Arrows represent incidence and emergence
directions. In the case of real data, if the measurements are taken
at different times, or over different regions, then one has to either:
(i) verify that the dataset corresponds to an homogeneous body, or
(ii) discriminate the heterogeneities. In this article, we only consider
the heterogeneous body by dividing the full dataset in two half homogeneous
bodies. \label{fig:Scheme-homogeous-VS-heterogeneous}}
\end{figure}

We present here two methods to decide if a dataset is homogeneous
or not (see Fig. \ref{fig:Scheme-homogeous-VS-heterogeneous}).

\subsubsection{Naive analysis of the estimated PDF\label{subsec:Stragety-for-separability-Naive-bayesian}}

One could expect that if the dataset is composed of two subsets with
different photometric parameters, then the estimated PDF would show
two modes around the true values. Then, discriminating the two sets
would be easy. One has to remind that this kind of test is not possible
with standard nonlinear least-squares optimization approach, since
only the best least-squares fit is searched (point estimation). 

\subsubsection{Chi-square-based strategy\label{subsec:Strategy-for-separability-chi2}}

One possible strategy consists of evaluating the plausibility of the
best obtained solution by analyzing the estimation residuals through
a chi-square analysis. This strategy can be summarized by the following
two steps:

\subsubsection*{1. Estimation of the best fit}

From the $N_{samp}$ samples obtained via the MCMC procedure, we consider
the best fit , noted $\tilde{r}_{i}$, that yields the minimum chi-square
value $\chi^{2}$, with:

\begin{equation}
\chi^{2}=\sum_{i}\frac{\left(r_{i}-\check{r_{\ell,i}}\right)^{2}}{\sigma_{i}^{2}},\label{eq:chi-square}
\end{equation}

which equivalently corresponds to the maximum likelihood value. Let
us note that the MCMC procedure explained in Section \ref{subsec:MCMC}
is used here in order to estimate the solution that yields the maximum
likelihood solution. Two remarks are now in order:
\begin{itemize}
\item First, since uniform prior distributions are used, the likelihood
$L$ and the posterior distribution $\sigma_{M}$ are proportional
in the definition domain of the parameters. Therefore, it makes sense
looking for the minimum of $L$ among the samples that are drawn according
to $\sigma_{M}$.
\item Second, although the MCMC algorithm is not strictly speaking an optimization
method, the optimization of $L$ could be addressed through non-linear
least-squares optimization methods such as the Levenberg-Marquardt
algorithm. However, the MCMC procedure asymptotically guarantees (as
$N_{samp}\rightarrow\infty)$ that the optimal solution is found,
whereas nonlinear least-squares optimization may converge to a local
(non-global) optimum.
\end{itemize}

\subsubsection*{2. Threshold on the $\chi^{2}$ value}

One can use the expected distribution of the $\chi^{2}$ value in
to test if the best fit case is compatible or not with the data and
the expected noise level (see section \ref{subsec:Noise-level}).
If we suppose that $r_{i}$ were correctly fitted by $\tilde{r}_{i}$,
then $\chi^{2}$ in Equation \ref{eq:chi-square} follows a chi-square
distribution with the degrees of freedom Nb. For a given probability
threshold (e.g., $\mathscr{P}$=5\%), the measured $\tilde{\chi}^{2}$
value is compared to the value $\chi_{lim}^{2}$ at which the chi-square
cumulative distribution function equals 1-$\mathscr{P}$. A measured
value larger than $\chi_{lim}^{2}$ means that there are less than
5\% chance that such a big value is due to noise only; then, we could
reject the model. An equivalent approach is to compute the value to
the cumulative probability of chi-square distribution $\mathscr{P}(\chi^{2}>\tilde{\chi}^{2})$
with N degrees of freedom, at a given $\tilde{\chi}^{2}$. If the
cumulative probability is larger than 5\%, we could reject the model.

\section{Synthetic tests}

We performed several synthetic observations in different conditions,
in order to propagate the uncertainties from observations into the
uncertainties on the Hapke parameters. First, the noise level rationale
is described. Then, the computation time improvement is discussed.
Third, the proposed efficiency $E$ is evaluated. Finally, the methodology
to decide if a dataset is homogeneous is evaluated.

\subsection{Noise level\label{subsec:Noise-level}}

For each test, a ``perfect'' model $\textrm{REFF}_{i}$ was computed
with known geometry $\Omega_{i}=(\theta_{0},\theta,g)_{i}$ using
Eq. \ref{eq:REFF}, and known parameters $m$. We model the uncertainties
on the measurements with additive Gaussian perturbations on $\textrm{REFF}_{i}$.
The standard deviation level $\sigma_{i}$ at geometry $i$ was set
to 10 \% of the observed reflectance $\textrm{REFF}_{i}$ in all the
numerical tests, except when specified. In case of very low reflectance
materials, the noise level is no more proportional to the $\textrm{REFF}$
measurements. We consider here that the minimum noise standard deviation
is 0.01 (as for instance for the instrument of IPAG, Grenoble as stated
in \citealt{Brissaud_SpectroGonio_AO2004}):

\begin{equation}
\sigma_{i}=\max(\frac{\textrm{REFF}_{i}}{10},0.01).\label{eq:NoiseLevel}
\end{equation}
The 10\% value may be overestimated for some spaceborne/laboratory
instrumental uncertainties. However, taking into account all error
sources (including atmosphere correction), a noise level of 10\% is
realistic \citep{Ceamanos_SurfacereflectanceMars_JoGRP2013,Fernando_SurfacereflectanceMars_JoGRP2013}.
For data obtained by the CRISM instrument (Compact Reconnaissance
Imaging Spectrometer for Mars), for instance, the reflectance error
at each geometry was estimated at $\sigma_{i}=r_{i}/50$ \citep{Fernando_SurfacereflectanceMars_JoGRP2013}.

In the following, regular synthetic reflectance values are generated
with additive, centered, Gaussian noise with variance $\sigma_{i}^{2}$.
Noise-free data are also considered. . Both synthetic noisy and noise-free
data are analyzed with an acceptable error in the likelihood (Eq.
\ref{eq:Likelyhood}) defined with Eq. \ref{eq:NoiseLevel}.

\subsection{Computation time improvement}

Both the former algorithm in \citealp{Schmidt_RealisticuncertaintiesHapke_I2015}
and the version described in Section \ref{subsec:MCMC} were implemented
on a personal computer equipped with a 2.9 GHz Intel Core i7 architecture.
Each algorithm was run several times with different random initializations.
Then, the number of samples was tuned such that the different estimates
of the mean of the parameters yield similar values, with at most 1\%
difference. In the first method, we considered that 1,000 \emph{accepted
samples} were necessary. We recall that the former algorithm (see
Algorithm 1 in \ref{sec:Appendix-Monte-Carlo-Markov}) suffered from
a very high rejection rate, that is, many iterations were necessary
in order to obtain one ''valid'' sample (typically less than 0.1\%
of acceptance rate, leading to typically 1,000,000 computed trials).
The required computation time was around 510 seconds in average. On
the contrary, with the new version, similar estimation is achieved
with 100,000\emph{ drawn samples} ( typically with 40\% acceptance
rate), which, in this case, corresponds to 100,000 iterations of Algorithm
2 in \ref{sec:Appendix-Monte-Carlo-Markov}. The required computation
time was around 13 seconds in average. Therefore, the computation
time was reduced by a factor of approximately 40 with the new implementation.

In the remainder of this article, all results were obtained by drawing
$N_{samp}=$100,000 samples, from which the first 5,000 were discarded,
that correspond to the burn-in regime of the MCMC procedure (initialization
effects). 

\subsection{Efficiency of EPF and BRDF}

Several authors proposed an optimized sampling of the EPF \citep{Schmidt_RealisticuncertaintiesHapke_I2015}
and the BRDF \citep{Souchon_experimentalstudyof_I2011} to get the
best information on the retrieved parameters. Such optimized sampling
is important for laboratory studies, when the recording time necessary
for sampling with many geometries may be prohibitive. It can also
be used to reduce the computation time of BRDF analysis by focusing
on the most important geometries, and to plan the best measurement
campaign in the case of spaceborne observations.

For one single geometry set, we evaluate $E$ by averaging 12 surface
types described by the parameters in Table \ref{tab:12HapkeParameters}
to cover the full range of possible photometries (high and low $\omega$;
narrow forward, intermediate and large backward surface $\{b,c\}$;
high and low $\bar{\theta}$).

\begin{table}
\begin{centering}
\begin{tabular}{|c|c|}
\hline 
 & Configurations\tabularnewline
\hline 
\hline 
$\omega$ (-) & 0.1, 0.7\tabularnewline
\hline 
$\bar{\theta}$ ($^{\circ}$) & 0.5, 25.0\tabularnewline
\hline 
$\{b\:;\:c\}$ (-) & \{0.1 ; 1.0\}, \{0.4 ; 0.4\}, \{0.8 ; 0.1\} \tabularnewline
\hline 
\end{tabular}
\par\end{centering}
\caption{Configurations of $\omega$, $\bar{\theta}$, $b$, $c$ used in the
estimation of the efficiency distance $E$ of BRDF. All 12 combinations
of those 4 parameter values were tested. \label{tab:12HapkeParameters}}
\end{table}

We tried geometries with the 23 sample directions proposed in \citealt{Souchon_experimentalstudyof_I2011}
(see Table \ref{tab:23souchon}); 23 random directions, sampled from
a uniform half-hemisphere (see Table \ref{tab:23random}); 23 directions
in the perpendicular plane, the worst case as defined by \citealt{Schmidt_RealisticuncertaintiesHapke_I2015}
(see Table \ref{tab:23worst-23best-64brdf}); 23 directions in the
principal plane (see Table \ref{tab:23worst-23best-64brdf}); 64 directions
to cover the full BRDF (see table \ref{tab:23worst-23best-64brdf}).
We choose to extensively test 23 directions in order to estimate the
effect of the geometry as proposed in \citealt{Souchon_experimentalstudyof_I2011},
used as a reference configuration. In comparison, we choose arbitrarily
64 directions to sample the full BRDF, 32 directions with 40$^{\circ}$
incidence, and 32 directions with 60$^{\circ}$ incidence. Each configuration
with 32 directions is defined by 4 emergence angles and 8 azimuth
angles in order to sample the full half-sphere.

Since phase angles close to 0 are tested, the opposition effect may
affect the result. We propose here two tests: the first one, with
realistic opposition effect, typical of the Moon ($B_{0}=1$, $h=0.1$)
\citep{Helfenstein_LunarOppositionEffect_I1997,Hillier_MultispectralPhotometryMoon_I1999};
the second one, ignoring the opposition effect ($B_{0}=0$).

\begin{table}
\begin{centering}
\begin{tabular}{|>{\centering}m{0.2cm}||>{\centering}m{0.2cm}|>{\centering}m{0.2cm}|>{\centering}m{0.2cm}|>{\centering}m{0.2cm}|>{\centering}m{0.2cm}|>{\centering}m{0.2cm}|>{\centering}m{0.2cm}|>{\centering}m{0.2cm}|>{\centering}m{0.2cm}|>{\centering}m{0.2cm}|>{\centering}m{0.2cm}|>{\centering}m{0.2cm}|>{\centering}m{0.2cm}|>{\centering}m{0.2cm}|>{\centering}m{0.2cm}|>{\centering}m{0.2cm}|>{\centering}m{0.2cm}|>{\centering}m{0.2cm}|>{\centering}m{0.2cm}|>{\centering}m{0.2cm}|>{\centering}m{0.2cm}|>{\centering}m{0.2cm}|>{\centering}m{0.2cm}|}
\hline 
 & \multicolumn{23}{c|}{23souchon}\tabularnewline
\hline 
$\theta_{0}$ & {\footnotesize{}10} & {\footnotesize{}10} & {\footnotesize{}30} & {\footnotesize{}30} & {\footnotesize{}30} & {\footnotesize{}30} & {\footnotesize{}30} & {\footnotesize{}50} & {\footnotesize{}50} & {\footnotesize{}50} & {\footnotesize{}30} & {\footnotesize{}30} & {\footnotesize{}30} & {\footnotesize{}50} & {\footnotesize{}50} & {\footnotesize{}50} & {\footnotesize{}50} & {\footnotesize{}60} & {\footnotesize{}60} & {\footnotesize{}60} & {\footnotesize{}45} & {\footnotesize{}55} & {\footnotesize{}55}\tabularnewline
\hline 
$\theta$ & {\footnotesize{}35} & {\footnotesize{}35 } & {\footnotesize{}60} & {\footnotesize{}0} & {\footnotesize{}20} & {\footnotesize{}40} & {\footnotesize{}60 } & {\footnotesize{}0} & {\footnotesize{}25} & {\footnotesize{}70} & {\footnotesize{}60} & {\footnotesize{}30} & {\footnotesize{}60} & {\footnotesize{}30} & {\footnotesize{}60} & {\footnotesize{}30} & {\footnotesize{}60} & {\footnotesize{}70} & {\footnotesize{}60} & {\footnotesize{}60} & {\footnotesize{}55} & {\footnotesize{}65} & {\footnotesize{}65}\tabularnewline
\hline 
$\varphi$  & {\footnotesize{}0} & {\footnotesize{}180} & {\footnotesize{}0} & {\footnotesize{}0} & {\footnotesize{}180} & {\footnotesize{}180} & {\footnotesize{}180 } & {\footnotesize{}0} & {\footnotesize{}0} & {\footnotesize{}180} & {\footnotesize{}45} & {\footnotesize{}135} & {\footnotesize{}135} & {\footnotesize{}45} & {\footnotesize{}45} & {\footnotesize{}135} & {\footnotesize{}135} & {\footnotesize{}180} & {\footnotesize{}45} & {\footnotesize{}135} & {\footnotesize{}90} & {\footnotesize{}45} & {\footnotesize{}135}\tabularnewline
\hline 
\end{tabular}
\par\end{centering}
\caption{Angular configuration, made of 23 directions, proposed by \citealp{Souchon_experimentalstudyof_I2011}.
$\theta_{0}$, $\theta$, and $\varphi$ represent incidence, emergence
and azimuth angles, respectively.\label{tab:23souchon}}
\end{table}

\begin{table}
\begin{centering}
\begin{tabular}{|>{\centering}p{0.2cm}||>{\centering}p{0.2cm}|>{\centering}p{0.2cm}|>{\centering}p{0.2cm}|>{\centering}p{0.2cm}|>{\centering}p{0.2cm}|>{\centering}p{0.2cm}|>{\centering}p{0.2cm}|>{\centering}p{0.2cm}|>{\centering}p{0.2cm}|>{\centering}p{0.2cm}|>{\centering}p{0.2cm}|>{\centering}p{0.2cm}|>{\centering}p{0.2cm}|>{\centering}p{0.2cm}|>{\centering}p{0.2cm}|>{\centering}p{0.2cm}|>{\centering}p{0.2cm}|>{\centering}p{0.2cm}|>{\centering}p{0.2cm}|>{\centering}p{0.2cm}|>{\centering}p{0.2cm}|>{\centering}p{0.2cm}|>{\centering}p{0.2cm}|}
\hline 
 & \multicolumn{23}{c|}{23random}\tabularnewline
\hline 
\hline 
$\theta_{0}$ & {\footnotesize{}14} & {\footnotesize{}35} & {\footnotesize{}68} & {\footnotesize{}76} & {\footnotesize{}23} & {\footnotesize{}78} & {\footnotesize{}46} & {\footnotesize{}18} & {\footnotesize{}55} & {\footnotesize{}45} & {\footnotesize{}74} & {\footnotesize{}30} & {\footnotesize{}59} & {\footnotesize{}78} & {\footnotesize{}39} & {\footnotesize{}4} & {\footnotesize{}23} & {\footnotesize{}46} & {\footnotesize{}58} & {\footnotesize{}64} & {\footnotesize{}52} & {\footnotesize{}45} & {\footnotesize{}25}\tabularnewline
\hline 
$\theta$ & {\footnotesize{}41} & {\footnotesize{}44} & {\footnotesize{}34} & {\footnotesize{}57} & {\footnotesize{}39} & {\footnotesize{}47} & {\footnotesize{}55} & {\footnotesize{}34} & {\footnotesize{}28} & {\footnotesize{}41} & {\footnotesize{}37} & {\footnotesize{}79} & {\footnotesize{}38} & {\footnotesize{}47} & {\footnotesize{}67} & {\footnotesize{}50} & {\footnotesize{}27} & {\footnotesize{}41} & {\footnotesize{}47} & {\footnotesize{}25} & {\footnotesize{}22} & {\footnotesize{}22} & {\footnotesize{}52}\tabularnewline
\hline 
$\varphi$  & {\footnotesize{}44} & {\footnotesize{}18} & {\footnotesize{}64} & {\footnotesize{}127} & {\footnotesize{}16} & {\footnotesize{}59} & {\footnotesize{}80} & {\footnotesize{}51} & {\footnotesize{}64} & {\footnotesize{}58} & {\footnotesize{}92 } & {\footnotesize{}109} & {\footnotesize{}58} & {\footnotesize{}61} & {\footnotesize{}101} & {\footnotesize{}53} & {\footnotesize{}43} & {\footnotesize{}85} & {\footnotesize{}66} & {\footnotesize{}79} & {\footnotesize{}64} & {\footnotesize{}31} & {\footnotesize{}27}\tabularnewline
\hline 
\end{tabular}
\par\end{centering}
\caption{Random angular configuration, made of 23 directions. The 23 directions
of incidence and emergence are sampled from an uniform distribution
of the upper hemisphere. $\theta_{0}$, $\theta$, and $\varphi$
represent incidence, emergence and azimuth angles, respectively.\label{tab:23random}}
\end{table}

\begin{table}
\begin{centering}
\begin{tabular}{|c|c|c|c|}
\hline 
 & 23worst & 23pplane & 64brdf\tabularnewline
\hline 
\hline 
$\theta_{0}$ (in $^{\circ}$) & $45$ & $75$ & $40$, $60$\tabularnewline
\hline 
$\theta$ (in $^{\circ}$) & {\small{}3.75:3.75:86.25} & 7.5:7.5:82.5, 0,82.5:7.5:7.5  & 10:20:70\tabularnewline
\hline 
$\varphi$ (in $^{\circ}$) & $90$ & $0^{11}$,0,$180^{11}$ & 0:45:315\tabularnewline
\hline 
\end{tabular}
\par\end{centering}
\caption{Definition of the angular configuration, made of 23 or 64 directions.
The worst are perpendicular to the principal plane and the best are
in the principal plane with incidence of 75$^{\circ}$. The notation
$\mathrm{X^{Y}}$ means that X is replicated Y times. The notation
X:Y:Z stands for X = first value, Y= step size, Z = last value. $\theta_{0}$,
$\theta$, and $\varphi$ represent incidence, emergence and azimuth
angles, respectively. \label{tab:23worst-23best-64brdf}}
\end{table}

Results are given in Tables \ref{tab:efficiency-geometry-set-with-opposition}
(taking into account the opposition effect) and \ref{tab:efficiency-geometry-set-without-opposition}
(without opposition effect). The best geometry set is obtained with
the 23 directions in the principal plane, which is generally even
better than using 64 BRDF directions, due to higher phase angle range.
As expected, choosing directions perpendicular to the principal plane
is the worst case. The geometries proposed by \citealp{Souchon_experimentalstudyof_I2011}
seem to be as efficient as random directions. Interestingly, the opposition
effect is not affecting significantly the results, as only a slight
improvement is noted when it is taken into account. This result shows
that the Bayesian method is able to extract information from the opposition
lobe separately from the other parameters, at least for these geometries.
This behavior was expected because the opposition lobe only occurs
for a limited number of geometries with small phase angles ($<20^{\circ}$).

\begin{table}
\begin{centering}
\begin{tabular}{|c|c|c|c|c|c|c|c|}
\hline 
 & {\scriptsize{}Global $E$} & {\scriptsize{}$E$ \#1} & {\scriptsize{}$E$ \#2} & {\scriptsize{}$E$ \#3} & {\scriptsize{}$E$ \#4} & {\scriptsize{}$E$ \#5} & {\scriptsize{}$E$ \#6}\tabularnewline
\hline 
{\scriptsize{}$\omega$} & {\scriptsize{}-} & {\scriptsize{}0.1} & {\scriptsize{}0.1} & {\scriptsize{}0.1} & {\scriptsize{}0.7} & {\scriptsize{}0.7} & {\scriptsize{}0.7}\tabularnewline
\hline 
{\scriptsize{}$\bar{\theta}$} & {\scriptsize{}-} & {\scriptsize{}0.5} & {\scriptsize{}0.5} & {\scriptsize{}0.5} & {\scriptsize{}0.5} & {\scriptsize{}0.5} & {\scriptsize{}0.5}\tabularnewline
\hline 
{\scriptsize{}$\{b\:;\:c\}$ } & {\scriptsize{}-} & {\scriptsize{}\{0.1 ; 1.0\}} & {\scriptsize{}\{0.4 ;0.4\}} & {\scriptsize{}\{0.8 ;0.1\} } & {\scriptsize{}\{0.1 ; 1.0\}} & {\scriptsize{}\{0.4 ;0.4\}} & {\scriptsize{}\{0.8 ;0.1\} }\tabularnewline
\hline 
\hline 
{\scriptsize{}23souchon} & {\scriptsize{}11.37} & {\scriptsize{}13.61 $\pm$ 0.38 } & {\scriptsize{}11.43 $\pm$ 0.32} & {\scriptsize{}12.41 $\pm$ 0.25 } & {\scriptsize{}11.82 $\pm$ 0.89 } & {\scriptsize{}8.99 $\pm$ 0.40} & {\scriptsize{}7.97 $\pm$ 0.25 }\tabularnewline
\hline 
{\scriptsize{}23random} & {\scriptsize{}10.91} & {\scriptsize{}13.24 $\pm$ 0.19 } & {\scriptsize{}11.47 $\pm$ 0.09 } & {\scriptsize{}11.99 $\pm$ 0.07 } & {\scriptsize{}11.55 $\pm$ 0.14 } & {\scriptsize{}8.78 $\pm$ 0.18 } & {\scriptsize{}7.08 $\pm$ 0.18 }\tabularnewline
\hline 
{\scriptsize{}23worst} & {\scriptsize{}14.21} & {\scriptsize{}15.59$\pm$ 0.25 } & {\scriptsize{}13.81 $\pm$ 0.10 } & {\scriptsize{}13.83 $\pm$ 0.05 } & {\scriptsize{}14.22 $\pm$ 0.24 } & {\scriptsize{}13.27 $\pm$ 0.16 } & {\scriptsize{}14.65 $\pm$ 0.27 }\tabularnewline
\hline 
{\scriptsize{}23pplane} & \textbf{\scriptsize{}8.79} & {\scriptsize{}12.62$\pm$ 0.51} & \textbf{\scriptsize{}7.37$\pm$ 0.32 } & \textbf{\scriptsize{}5.26 $\pm$ 0.17 } & \textbf{\scriptsize{}10.02 $\pm$ 0.21 } & \textbf{\scriptsize{}7.03 $\pm$ 0.17} & \textbf{\scriptsize{}4.38 $\pm$ 0.14 }\tabularnewline
\hline 
{\scriptsize{}64brdf} & {\scriptsize{}9.26} & \textbf{\scriptsize{}11.60 $\pm$ 0.23 } & {\scriptsize{}9.11 $\pm$ 0.20} & {\scriptsize{}9.31 $\pm$ 0.13 } & {\scriptsize{}10.80 $\pm$ 0.31 } & {\scriptsize{}7.97 $\pm$ 0.10 } & {\scriptsize{}4.89 $\pm$ 0.18}\tabularnewline
\hline 
\end{tabular}
\par\end{centering}
\begin{tabular}{|c|c|c|c|c|c|c|}
\hline 
 & {\scriptsize{}$E$ \#7} & {\scriptsize{}$E$ \#8} & {\scriptsize{}$E$ \#9} & {\scriptsize{}$E$ \#10} & {\scriptsize{}$E$ \#11} & {\scriptsize{}$E$ \#12}\tabularnewline
\hline 
{\scriptsize{}$\omega$} & {\scriptsize{}0.1} & {\scriptsize{}0.1} & {\scriptsize{}0.1} & {\scriptsize{}0.7} & {\scriptsize{}0.7} & {\scriptsize{}0.7}\tabularnewline
\hline 
{\scriptsize{}$\bar{\theta}$} & {\scriptsize{}25.0} & {\scriptsize{}25.0} & {\scriptsize{}25.0} & {\scriptsize{}25.0} & {\scriptsize{}25.0} & {\scriptsize{}25.0}\tabularnewline
\hline 
{\scriptsize{}$\{b\:;\:c\}$ } & {\scriptsize{}\{0.1 ; 1.0\}} & {\scriptsize{}\{0.4 ;0.4\}} & {\scriptsize{}\{0.8 ;0.1\} } & {\scriptsize{}\{0.1 ; 1.0\}} & {\scriptsize{}\{0.4 ;0.4\}} & {\scriptsize{}\{0.8 ;0.1\} }\tabularnewline
\hline 
\hline 
{\scriptsize{}23souchon} & {\scriptsize{}13.48 $\pm$ 0.20} & {\scriptsize{}12.59 $\pm$ 0.21 } & {\scriptsize{}13.78 $\pm$ 0.15 } & {\scriptsize{}11.88 $\pm$ 0.38 } & {\scriptsize{}9.66 $\pm$ 0.36} & {\scriptsize{}8.86 $\pm$ 0.55}\tabularnewline
\hline 
{\scriptsize{}23random} & {\scriptsize{}13.09 $\pm$ 0.14} & {\scriptsize{}11.95 $\pm$ 0.04 } & {\scriptsize{}13.10 $\pm$ 0.05 } & {\scriptsize{}11.75 $\pm$ 0.28 } & {\scriptsize{}9.62 $\pm$ 0.09 } & {\scriptsize{}7.26 $\pm$ 0.32}\tabularnewline
\hline 
{\scriptsize{}23worst} & {\scriptsize{}15.30 $\pm$ 0.16} & {\scriptsize{}13.66 $\pm$ 0.07} & {\scriptsize{}13.71 $\pm$ 0.04 } & {\scriptsize{}15.06 $\pm$ 0.28} & {\scriptsize{}13.48 $\pm$ 0.13} & {\scriptsize{}13.99 $\pm$ 0.10}\tabularnewline
\hline 
{\scriptsize{}23pplane} & {\scriptsize{}13.85}\textbf{\scriptsize{} }{\scriptsize{}$\pm$ 0.29} & {\scriptsize{}11.39 $\pm$ 0.22} & \textbf{\scriptsize{}8.76 $\pm$ 0.32} & {\scriptsize{}12.02 $\pm$ 0.42 } & \textbf{\scriptsize{}8.39 $\pm$ 0.32 } & \textbf{\scriptsize{}4.42 $\pm$ 0.31}\tabularnewline
\hline 
{\scriptsize{}64brdf} & \textbf{\scriptsize{}12.93 $\pm$ 0.18 } & \textbf{\scriptsize{}10.54 $\pm$ 0.13}{\scriptsize{} } & {\scriptsize{}10.89 $\pm$ 0.17 } & \textbf{\scriptsize{}10.64 $\pm$ 0.30} & \textbf{\scriptsize{}8.28 $\pm$ 0.24 } & \textbf{\scriptsize{}4.22 $\pm$ 0.30}\tabularnewline
\hline 
\end{tabular}

\caption{Efficiency distance $E$ of each angular configuration set used to
estimate the true Hapke parameters, on 12 photometric surfaces (from
\#1 to \#12), taking into account the opposition effect. Average values
and standard deviations are computed over 10 independent experiments.
The global efficiency is computed as the average over the 12 photometric
surfaces. Best results for each surface are highlighted in bold. \label{tab:efficiency-geometry-set-with-opposition}}
\end{table}

\begin{table}
\begin{centering}
\begin{tabular}{|c|c|c|c|c|c|c|c|}
\hline 
 & {\scriptsize{}Global $E$} & {\scriptsize{}$E$ \#1} & {\scriptsize{}$E$ \#2} & {\scriptsize{}$E$ \#3} & {\scriptsize{}$E$ \#4} & {\scriptsize{}$E$ \#5} & {\scriptsize{}$E$ \#6}\tabularnewline
\hline 
{\scriptsize{}$\omega$} & {\scriptsize{}-} & {\scriptsize{}0.1} & {\scriptsize{}0.1} & {\scriptsize{}0.1} & {\scriptsize{}0.7} & {\scriptsize{}0.7} & {\scriptsize{}0.7}\tabularnewline
\hline 
{\scriptsize{}$\bar{\theta}$} & {\scriptsize{}-} & {\scriptsize{}0.5} & {\scriptsize{}0.5} & {\scriptsize{}0.5} & {\scriptsize{}0.5} & {\scriptsize{}0.5} & {\scriptsize{}0.5}\tabularnewline
\hline 
{\scriptsize{}$\{b\:;\:c\}$ } & {\scriptsize{}-} & {\scriptsize{}\{0.1 ; 1.0\}} & {\scriptsize{}\{0.4 ;0.4\}} & {\scriptsize{}\{0.8 ;0.1\} } & {\scriptsize{}\{0.1 ; 1.0\}} & {\scriptsize{}\{0.4 ;0.4\}} & {\scriptsize{}\{0.8 ;0.1\} }\tabularnewline
\hline 
\hline 
{\scriptsize{}23souchon} & {\scriptsize{}11.22} & {\scriptsize{}13.60 $\pm$ 0.14 } & {\scriptsize{}11.41 $\pm$ 0.17} & {\scriptsize{}12.43 $\pm$ 0.07 } & {\scriptsize{}11.59 $\pm$ 0.15 } & {\scriptsize{}8.72 $\pm$ 0.08} & {\scriptsize{}7.43 $\pm$ 0.19 }\tabularnewline
\hline 
{\scriptsize{}23random} & {\scriptsize{}11.00} & {\scriptsize{}13.30 $\pm$ 0.40 } & {\scriptsize{}11.35 $\pm$ 0.26 } & {\scriptsize{}11.90 $\pm$ 0.52 } & {\scriptsize{}11.46 $\pm$ 0.37 } & {\scriptsize{}8.73 $\pm$ 0.22 } & {\scriptsize{}7.43 $\pm$ 0.24 }\tabularnewline
\hline 
{\scriptsize{}23worst} & {\scriptsize{}14.30} & {\scriptsize{}15.92$\pm$ 0.80 } & {\scriptsize{}13.80 $\pm$ 0.07 } & {\scriptsize{}13.77 $\pm$ 0.08 } & {\scriptsize{}14.84 $\pm$ 1.30 } & {\scriptsize{}13.23 $\pm$ 0.17 } & {\scriptsize{}14.64 $\pm$ 0.39 }\tabularnewline
\hline 
{\scriptsize{}23pplane} & \textbf{\scriptsize{}8.31} & \textbf{\scriptsize{}10.46}{\scriptsize{}$\pm$}\textbf{\scriptsize{}
0.81} & \textbf{\scriptsize{}6.60$\pm$ 0.60 } & \textbf{\scriptsize{}4.41 $\pm$ 0.26 } & \textbf{\scriptsize{}9.69 $\pm$ 0.65 } & \textbf{\scriptsize{}6.75 $\pm$ 0.41} & \textbf{\scriptsize{}4.12 $\pm$ 0.27 }\tabularnewline
\hline 
{\scriptsize{}64brdf} & {\scriptsize{}9.14} & {\scriptsize{}11.35 $\pm$ 0.28 } & {\scriptsize{}8.94 $\pm$ 0.29} & {\scriptsize{}9.56 $\pm$ 0.46 } & \textbf{\scriptsize{}10.29 $\pm$ 0.37}{\scriptsize{} } & {\scriptsize{}7.66 $\pm$ 0.27 } & {\scriptsize{}4.99 $\pm$ 0.40}\tabularnewline
\hline 
\end{tabular}
\par\end{centering}
\begin{tabular}{|c|c|c|c|c|c|c|}
\hline 
 & {\scriptsize{}$E$ \#7} & {\scriptsize{}$E$ \#8} & {\scriptsize{}$E$ \#9} & {\scriptsize{}$E$ \#10} & {\scriptsize{}$E$ \#11} & {\scriptsize{}$E$ \#12}\tabularnewline
\hline 
{\scriptsize{}$\omega$} & {\scriptsize{}0.1} & {\scriptsize{}0.1} & {\scriptsize{}0.1} & {\scriptsize{}0.7} & {\scriptsize{}0.7} & {\scriptsize{}0.7}\tabularnewline
\hline 
{\scriptsize{}$\bar{\theta}$} & {\scriptsize{}25.0} & {\scriptsize{}25.0} & {\scriptsize{}25.0} & {\scriptsize{}25.0} & {\scriptsize{}25.0} & {\scriptsize{}25.0}\tabularnewline
\hline 
{\scriptsize{}$\{b\:;\:c\}$ } & {\scriptsize{}\{0.1 ; 1.0\}} & {\scriptsize{}\{0.4 ;0.4\}} & {\scriptsize{}\{0.8 ;0.1\} } & {\scriptsize{}\{0.1 ; 1.0\}} & {\scriptsize{}\{0.4 ;0.4\}} & {\scriptsize{}\{0.8 ;0.1\} }\tabularnewline
\hline 
\hline 
{\scriptsize{}23souchon} & {\scriptsize{}13.39 $\pm$ 0.15 } & {\scriptsize{}12.48 $\pm$ 0.06 } & {\scriptsize{}13.82 $\pm$ 0.07 } & {\scriptsize{}11.68 $\pm$ 0.17 } & {\scriptsize{}9.36 $\pm$ 0.24} & {\scriptsize{}8.73 $\pm$ 0.18}\tabularnewline
\hline 
{\scriptsize{}23random} & {\scriptsize{}13.15 $\pm$ 0.46 } & {\scriptsize{}11.89 $\pm$ 0.18 } & {\scriptsize{}13.36 $\pm$ 0.27 } & {\scriptsize{}11.79 $\pm$ 0.33 } & {\scriptsize{}10.11 $\pm$ 0.38 } & {\scriptsize{}7.48 $\pm$ 0.32}\tabularnewline
\hline 
{\scriptsize{}23worst} & {\scriptsize{}15.38 $\pm$ 0.17 } & {\scriptsize{}13.67 $\pm$ 0.08} & {\scriptsize{}13.68 $\pm$ 0.06 } & {\scriptsize{}15.30 $\pm$ 0.47} & {\scriptsize{}13.49 $\pm$ 0.15} & {\scriptsize{}13.89 $\pm$ 0.14}\tabularnewline
\hline 
{\scriptsize{}23pplane} & \textbf{\scriptsize{}13.23 $\pm$ 0.68 } & \textbf{\scriptsize{}10.86 $\pm$ 0.33} & \textbf{\scriptsize{}8.29 $\pm$ 0.21} & {\scriptsize{}11.97 $\pm$ 0.56 } & {\scriptsize{}8.93 $\pm$ 0.53 } & \textbf{\scriptsize{}4.41 $\pm$ 0.97}\tabularnewline
\hline 
{\scriptsize{}64brdf} & \textbf{\scriptsize{}12.58 $\pm$ 0.56 } & \textbf{\scriptsize{}10.55 $\pm$ 0.44}{\scriptsize{} } & {\scriptsize{}10.93 $\pm$ 0.43 } & \textbf{\scriptsize{}11.04 $\pm$ 0.55} & \textbf{\scriptsize{}7.98 $\pm$ 0.27 } & \textbf{\scriptsize{}3.79 $\pm$ 0.45}\tabularnewline
\hline 
\end{tabular}

\caption{Similar results to those in Figure \ref{tab:efficiency-geometry-set-with-opposition},
but without taking into account the opposition effect.\label{tab:efficiency-geometry-set-without-opposition}}
\end{table}

\begin{figure}
\includegraphics[width=1\textwidth]{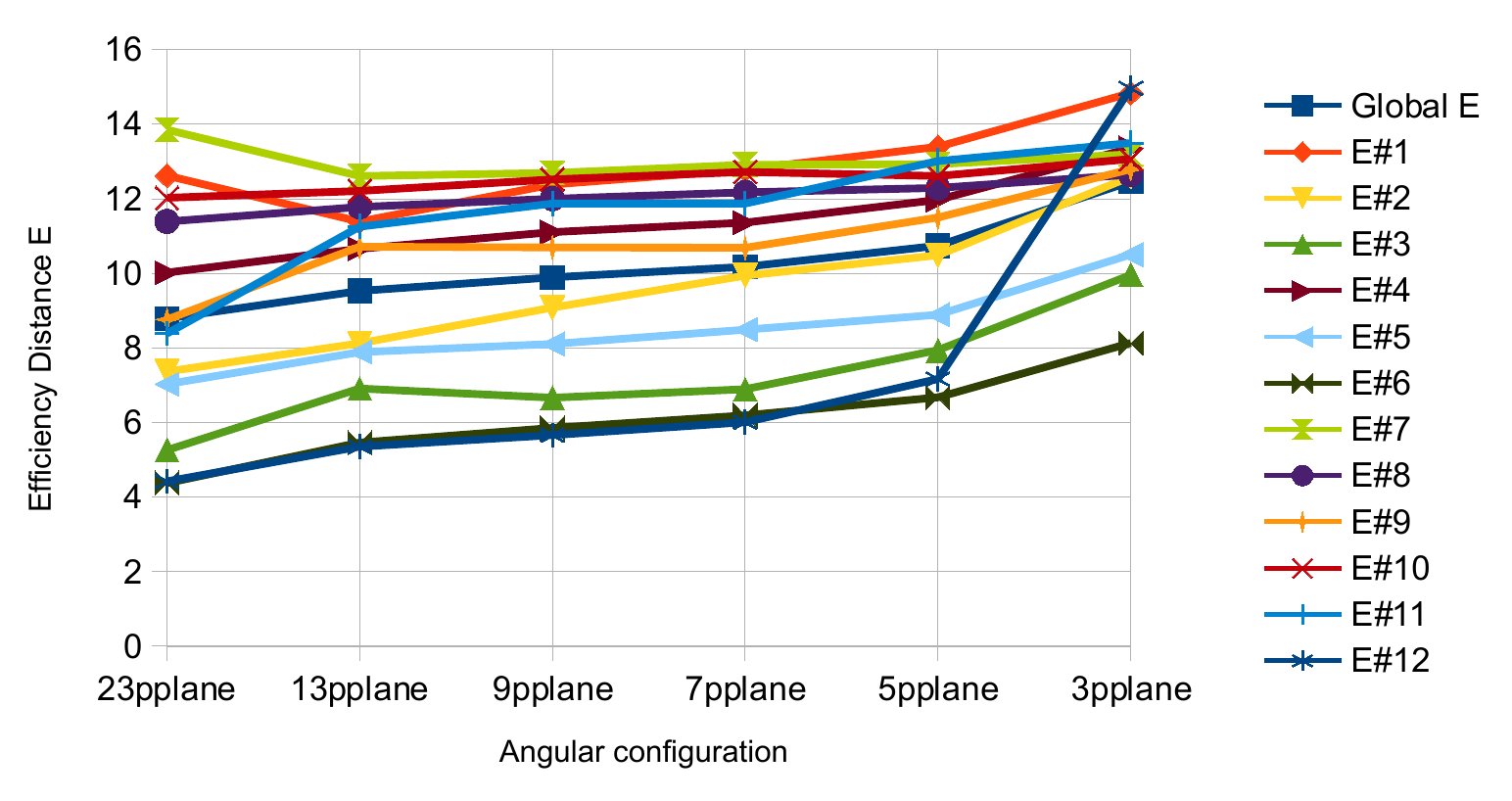}

\caption{Efficiency distance $E$ of each angular configuration set used to
estimate the true Hapke parameters, on 12 photometric surfaces (from
\#1 to \#12), taking into account the opposition effect, for decreasing
number of geometries with the same phase range. 23pplane is the configuration
from Table \ref{tab:23worst-23best-64brdf}. For all other angular
configuration sets, nadir and extreme emergence angles are kept. Remaining
angles are decimated to study the effect of loosing information in
the dataset. \label{fig:Efficiency_pplane_nbConfig}}
\end{figure}

Another experiment is proposed in order to study the behavior of the
efficiency distance $E$ in the principal plane, but with a decreasing
number of geometries. This kind of experiment was already tested in
\citealt{Schmidt_RealisticuncertaintiesHapke_I2015}, which demonstrated
the effect of the azimuth in the Emission Phase Function (EPF) configuration
and the effect of the phase angle range. Here, we aim at quantifying
the effect of the number of angular configurations in the same phase
angle range. We used the 23pplane values (see table \ref{tab:23worst-23best-64brdf})
as a reference and then we decimate the number of angular configuration
to 13, 9, 7, 5 and 3. In all cases, nadir (0\textdegree ) and the
two extreme (82.5\textdegree ) emergence angles are kept. The intermediate
directions are regularly sampled among the possible values. Results
are shown in Figure \ref{fig:Efficiency_pplane_nbConfig}. From 23
to 5 geometries, the efficiency distance $E$ increases slowly due
to loss of information in the dataset, in average (global $E$), but
also for the 12 photometric surfaces (from \#1 to \#12). We notice
that using 3 angular configurations dramatically decreases the estimation
performance compared with 5 configurations. Thus, we can conclude
that 5 directions is the minimum number of angular configurations
in the best situation (principal plane) in order to expect well constrained
photometric parameters.

\begin{figure}
\includegraphics[width=1\textwidth]{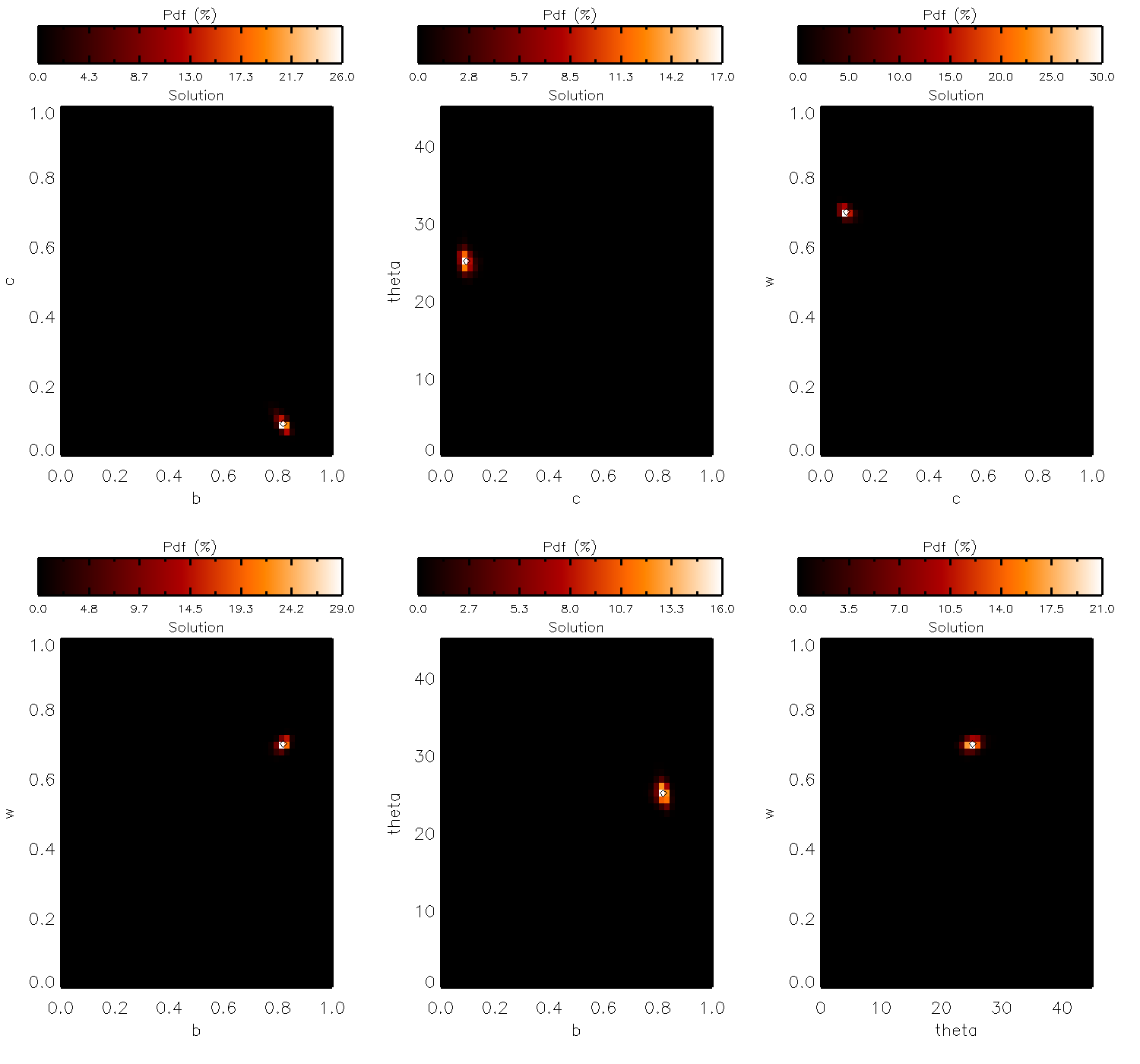}

\caption{Example of one of the best results ($E=3.96$), with 2D-histograms
of the generated samples: 23pplane geometry and configuration \#12
(see Table \ref{tab:efficiency-geometry-set-without-opposition}).\label{fig:Example-of-one-best}}
\end{figure}

\begin{figure}
\includegraphics[width=1\textwidth]{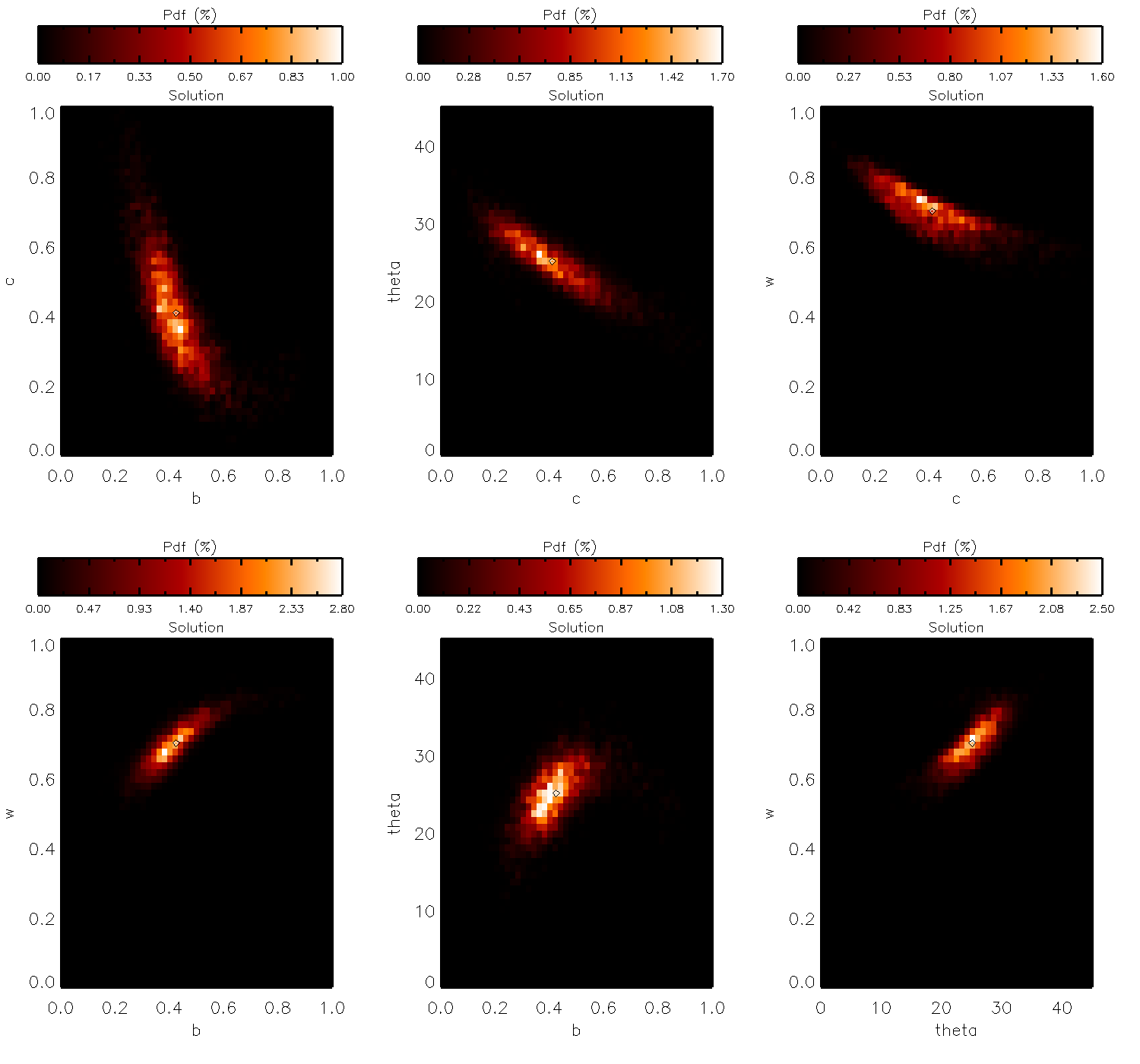}

\caption{Example of one average result ($E=9.74$), with 2D-histograms of the
generated samples: 23souchon geometry and configuration \#11 (see
Table \ref{tab:efficiency-geometry-set-without-opposition}). \label{fig:Example-of-an-average}}
\end{figure}

\begin{figure}
\includegraphics[width=1\textwidth]{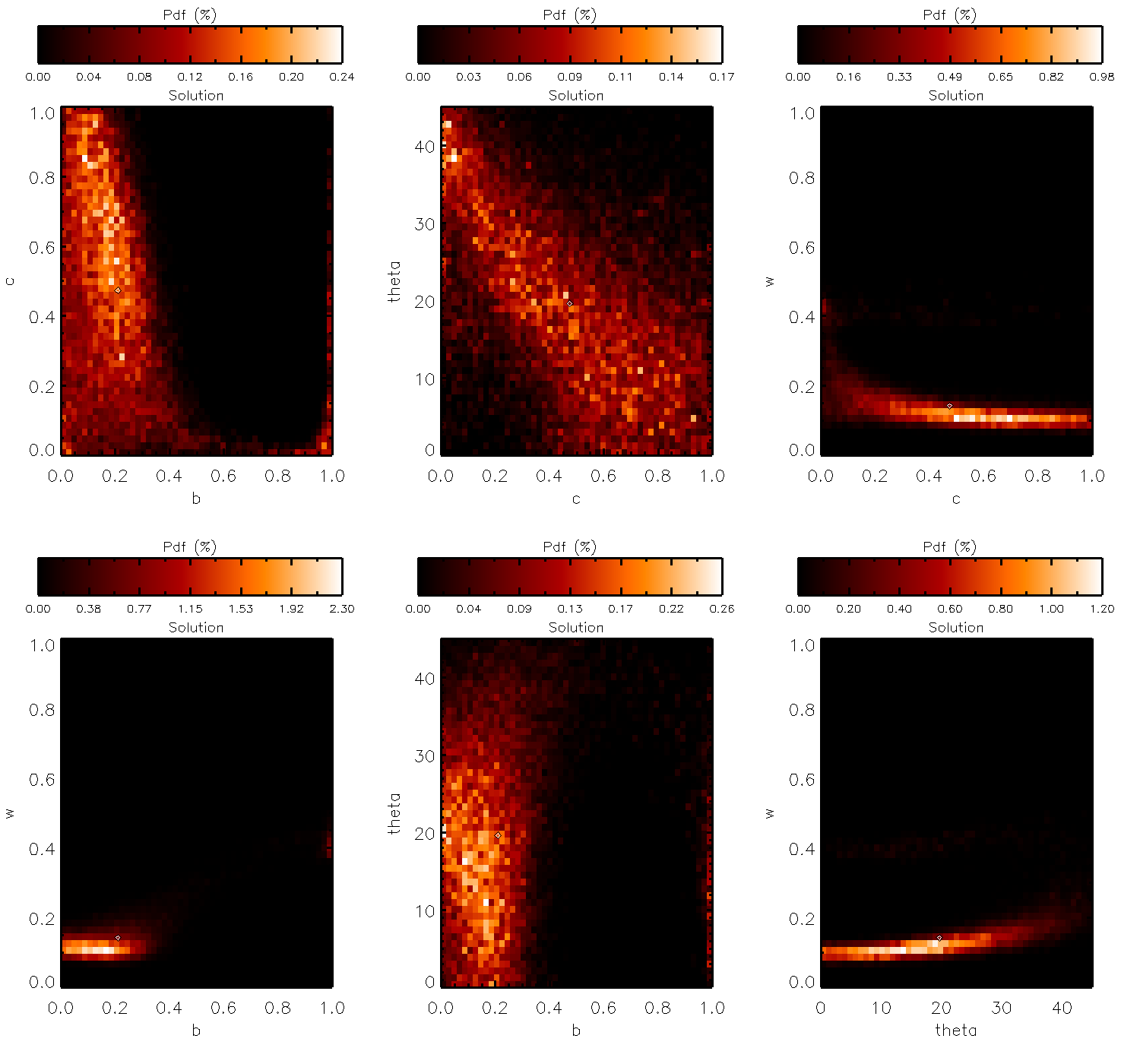}

\caption{Example of one of the worst result ($E=16.19$), with 2D-histograms
of the generated samples: 23worst geometry and configuration \#1 (see
Table \ref{tab:efficiency-geometry-set-without-opposition}).\label{fig:Example-of-one-worst}}
\end{figure}

In order to have a better understanding of the parameter $E$, Figures
\ref{fig:Example-of-one-best} to \ref{fig:Example-of-one-worst}
represent cases close to the best, the worst and the average situations,
in form of 2D histograms. The best cases ($E\sim4$), shown in Fig.
\ref{fig:Example-of-one-best}, display a very localized result, around
the true value. The worst case ($E\sim16$) presented in Fig. \ref{fig:Example-of-one-worst},
indicates that the solution represents a large fraction of the possible
domain. An example close to the average efficiency for a standard
geometry ($E\sim10$), represented in Fig. \ref{fig:Example-of-an-average},
shows an intermediate situation.

\subsection{Heterogeneity of two unknown BRDFs}

\begin{figure}
\hfill{}\includegraphics[width=0.4\textwidth]{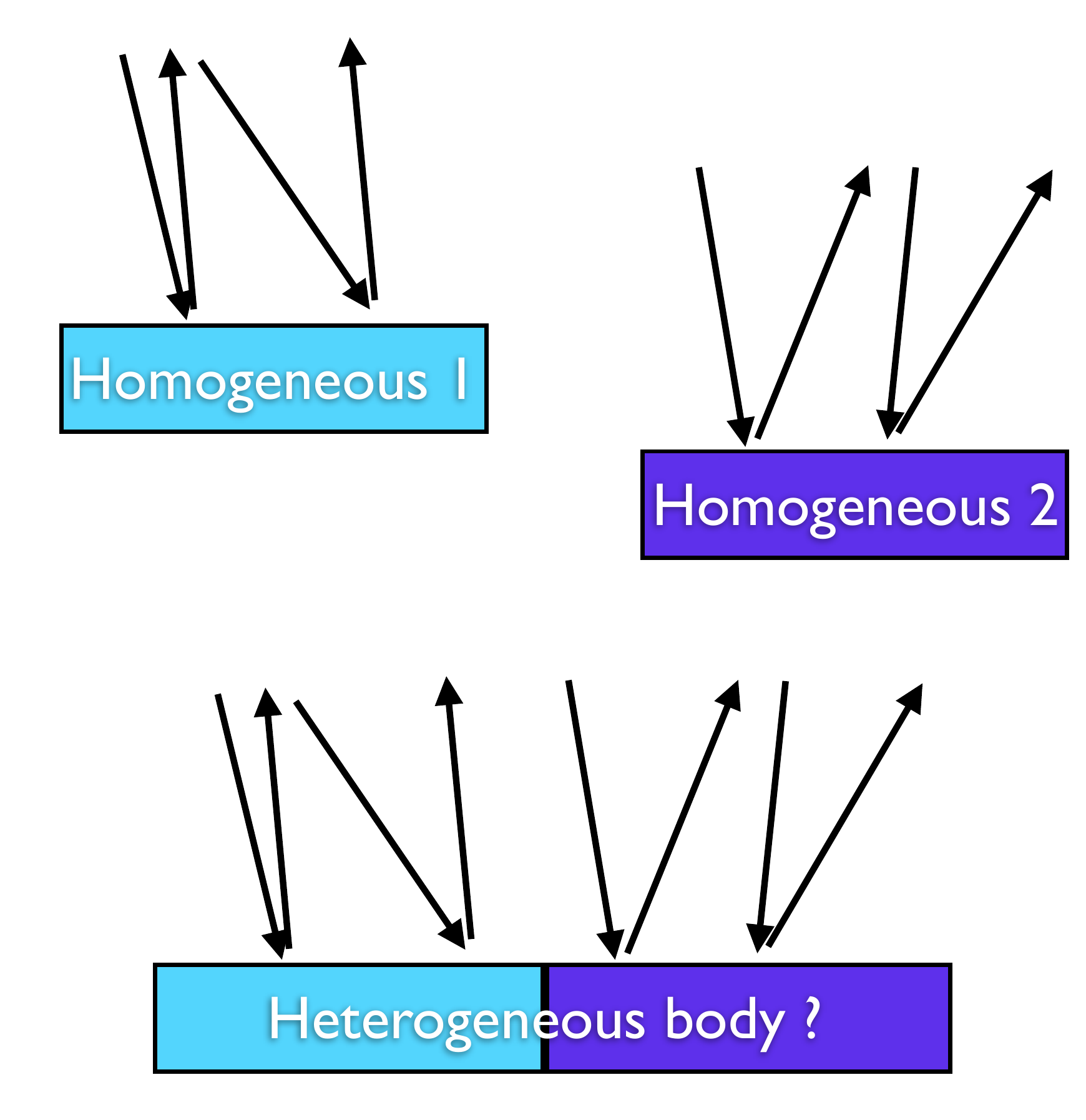}\hfill{}

\caption{Sketch of the synthetic heterogeneous body dataset building. Arrows
represent the observation geometries. Half of the dataset is taken
from one surface type, the other half from another surface type. \label{fig:BuildingHeterogeneousDataset}}
\end{figure}

In this section, we generate two synthetic datasets, with distinct
photometric properties in order to test our separability approach
(see Section \ref{subsec:Strategy-for-separability}). Figure \ref{fig:BuildingHeterogeneousDataset}
represents a sketch of how we built the synthetic heterogeneous body
dataset. We built a synthetic observation: 50 observations on the
region 1 of a planet and 50 other observations on region 2. We can
imagine a spacecraft/telescope observing a planetary body, with 50
observations on one side (region 1), and 50 on other side (region
2). All 100 observations have a different geometry. The planetary
body has two geological terrains: region 1 on one side, region 2 on
the other side. The question is: can we put all observations together
and determine a good fit? Can we find a way to argue that the dataset
is heterogeneous?

An equivalent experiment can also be done: the interpretation of 50
images, both observing region 1 and region 2. The analyst picks for
each image 1 pixel of region 1 and 1 pixel of region 2. The final
dataset consists of 50 geometries on region 1 and 50 geometries on
region 2, but with 50 identical geometries. The results of this second
experiment is present in the supplementary material of this article
and is leading to similar conclusions.

In the first section, we test our approach on some toy problems, for
pedagogical purposes. We compare the results obtained by considering
each dataset separately, and by considering the two datasets together.
We demonstrate that the naive analysis of the solution PDF does not
yield satisfactory results. In the second section, we propose to evaluate
the chi-square-based method in a more general scope. 

\subsubsection{Separability on some examples\label{subsec:Separability-on-some-examples}}

\begin{figure}
\includegraphics[clip,width=1.0\textwidth]{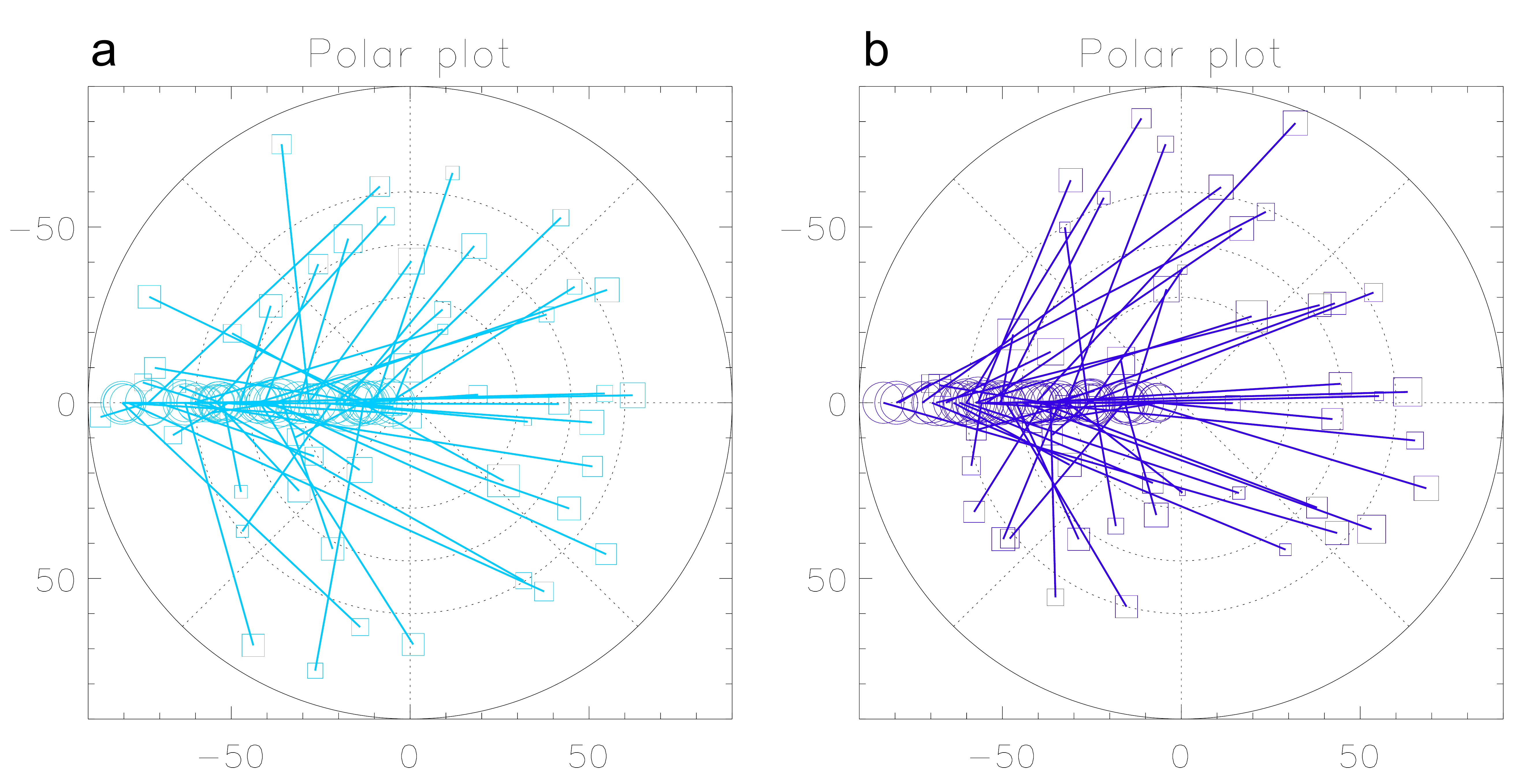}\centering\caption{Polar plot representing 100 random geometries used in the experiments
of Section \ref{subsec:Separability-on-some-examples}. Each geometry
is represented by its incidence (circle) and emergence (square) angles,
linked by a straight line. (a) The 50 geometries of brdf1/brdf3/brdf5
are in light blue and (b) 50 geometries of brdf2/brdf4/brdf6 are in dark
blue (please see the online version for color figures).\label{fig:Polarplot-100random}}
\end{figure}

We consider the following tests: 100 directions are randomly generated
uniformly in the upper hemisphere (see fig. \ref{fig:Polarplot-100random}).
The 100 directions are divided into two groups (region 1 and region
2) with only one photometric parameter varying between them. Table
\ref{tab:brf_3cases} summarizes the three cases we adopt in order
to compare the effect of: phase function (brdf1 vs brdf2), single
scattering albedo (brdf3 vs brdf4), macroscopic roughness (brdf5 vs
brdf6). We consider both noise-free and noisy data with 10\% noise,
as explained in Section \ref{subsec:Noise-level}.

\begin{figure}
\includegraphics[bb=0bp 0bp 604bp 861bp,clip,width=1\textwidth]{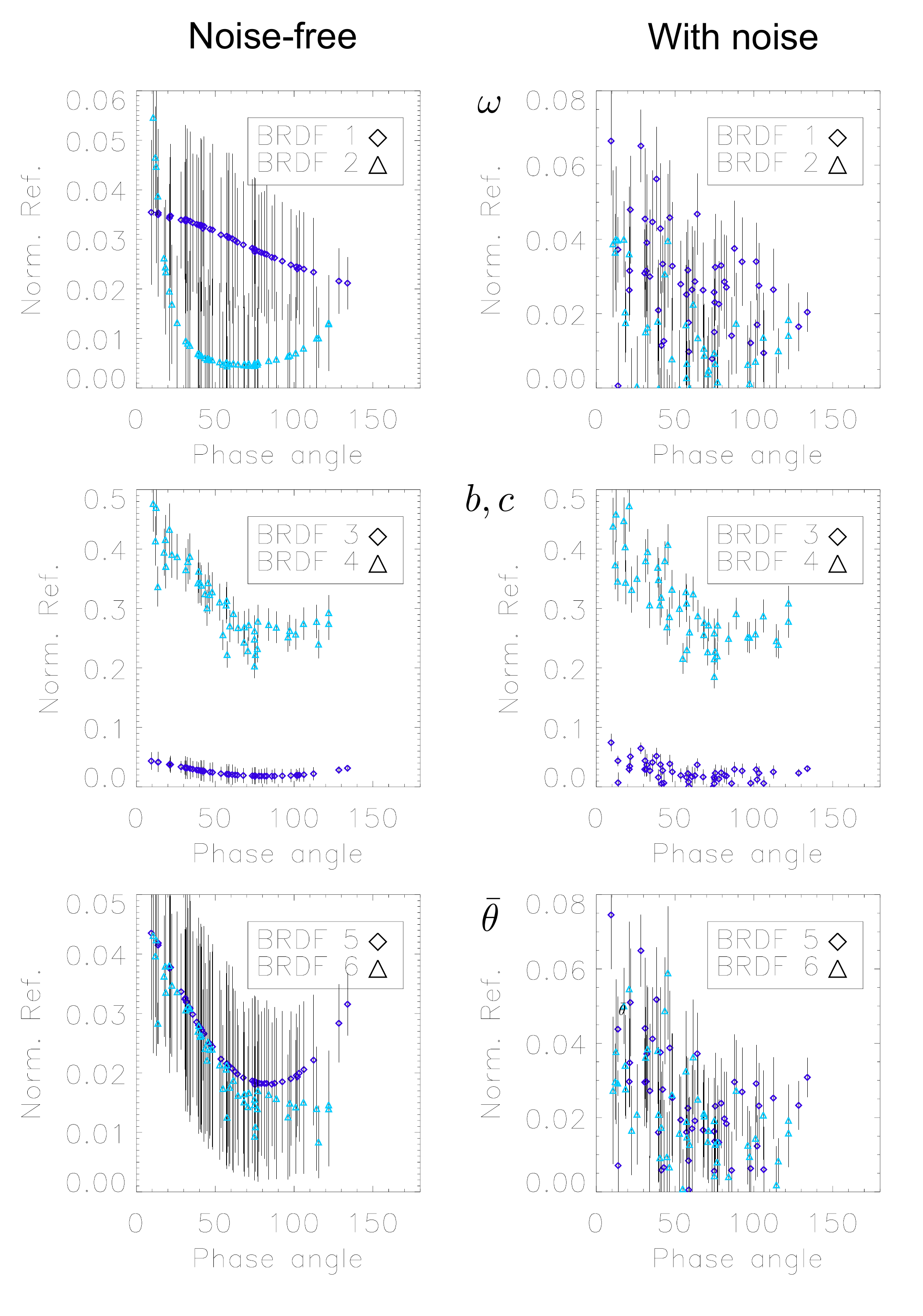}

\caption{Normalized reflectance without noise (left) and with noise (right)
as a function of the phase angle for the 100 random geometries in
Figure \ref{fig:Polarplot-100random}. The 50 geometries of brdf1,3,5
are in light blue and 50 geometries of brdf2,4,6 are in dark blue
(please see the online version for color figures). Top (respectively,
middle and bottom) panels represent two different values in b/c (respectively,
$\omega$, and $\bar{\theta}$). See Table \ref{tab:brf_3cases} for
the corresponding numerical values. The symbols represent the two
subsets composing the heterogeneous data (see Fig. \ref{fig:BuildingHeterogeneousDataset}).
All 100 geometries (50 for each subset) are represented in these figures.
Errors bars are defined in Eq. \ref{eq:NoiseLevel} and used as values
for $\sigma_{i}$ in the likelihood (see Eq. \ref{eq:Likelyhood}).
The noise-free case contains no random noise. \label{fig:brdf135-brdf246_Refnorm-RefnormNOISE_VS_Phase}}
\end{figure}

The normalized reflectance (eq. \ref{eq:REFF-normalized}) is represented
in Figure \ref{fig:brdf135-brdf246_Refnorm-RefnormNOISE_VS_Phase}
as a function of the phase angle, in the noise-free (left) and noisy
(right) cases. In the ideal noise-free case, one can easily separate
by eye two extreme phase functions (brdf1 vs. brdf2), or single scattering
albedo (brdf3 vs brdf4). The two different macroscopic roughnesses,
however, are difficult to distinguish (brdf5 vs brdf6). In the presence
of noise, only the single scattering albedo case (brdf3 vs brdf4)
can be deciphered by eye. In the following, we study if it is possible
to distinguish between those three cases using the two proposed approaches.
Note, of course, that in more realistic cases, a mixture of two surfaces
generally shows differences in \emph{all} photometric parameters (not
only one). This case is considered in the next section. This section
simply aims to illustrate and validate a possible approach on particular
cases. 

\begin{table}
\begin{centering}
\begin{tabular}{|c||c|c||c|c||c|c|}
\hline 
 & 50brdf1 & 50brdf2 & 50brdf3 & 50brdf4 & 50brdf5 & 50brdf6\tabularnewline
\hline 
\hline 
$\omega$ & 0.1  & 0.1 & \textbf{0.1}  & \textbf{0.7} & 0.1  & 0.1\tabularnewline
\hline 
$\bar{\theta}$ & 0.5$^{\circ}$ & 0.5$^{\circ}$ & 0.5$^{\circ}$ & 0.5$^{\circ}$ & \textbf{0.5$^{\circ}$} & \textbf{25}\emph{$^{\circ}$}\tabularnewline
\hline 
$b$ & \textbf{0.1} & \textbf{0.8} & 0.4 & 0.4 & 0.4 & 0.4\tabularnewline
\hline 
$c$ & \textbf{1.0} & \textbf{0.1} & 0.4 & 0.4 & 0.4 & 0.4\tabularnewline
\hline 
\end{tabular}
\par\end{centering}
\caption{Set of photometric parameters used in the separability experiments
(section \ref{subsec:Separability-on-some-examples}) \label{tab:brf_3cases}}
\end{table}

\subsubsection*{Naive approach}

\begin{figure}
\includegraphics[width=1\textwidth]{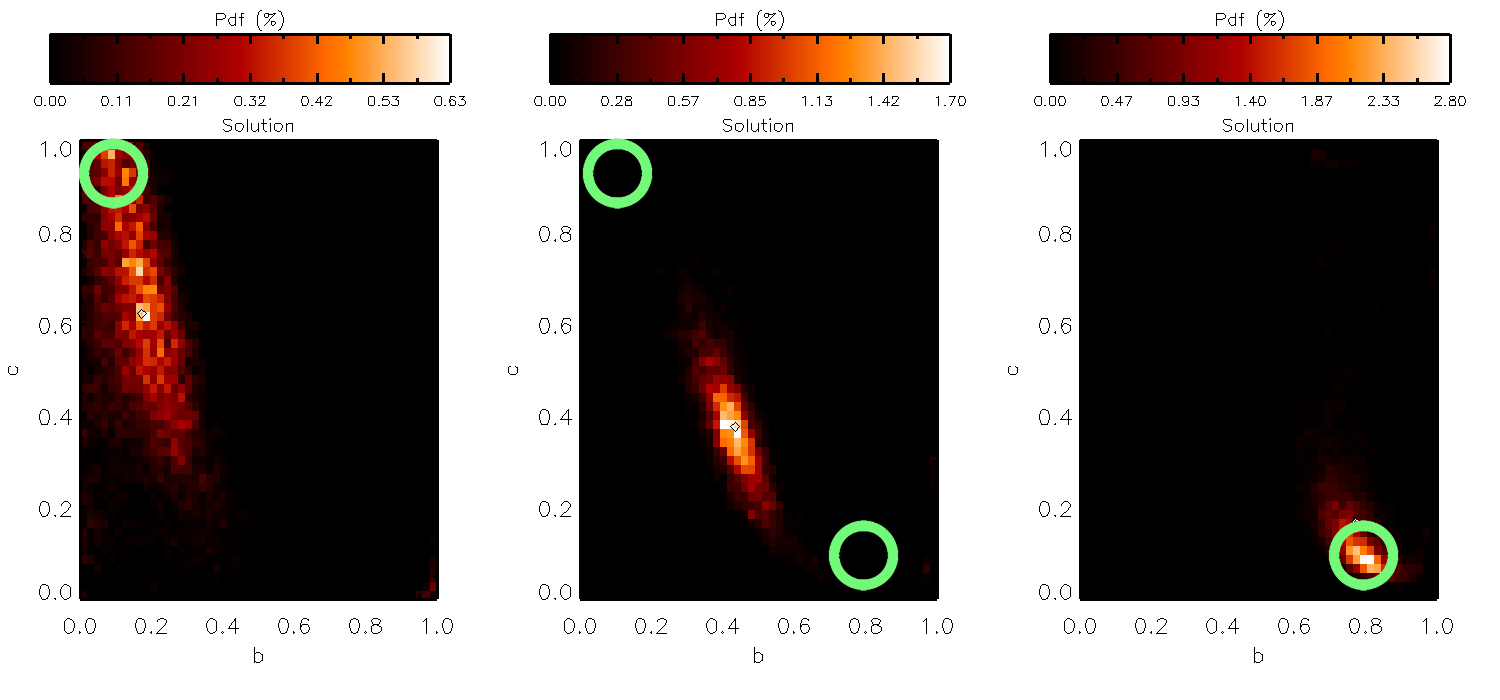}

\caption{Estimated PDF on the 2D histogram ($c$ vs. $b)$ obtained from 50
geometries of brdf1 (left) and brdf2 (right) and from the mixture
(center) corresponding to figure \ref{fig:brdf135-brdf246_Refnorm-RefnormNOISE_VS_Phase}
top right (noisy case). Green circles locate the true values. The
concatenation of two photometric situations (broad backward for brdf1
and narrow forward for brdf2) leads to a misinterpretation of an intermediate
photometric situation. \label{fig:brdf1-brdf2_b_VS_c}}
\end{figure}

\begin{figure}
\hfill{}\includegraphics[width=1\textwidth]{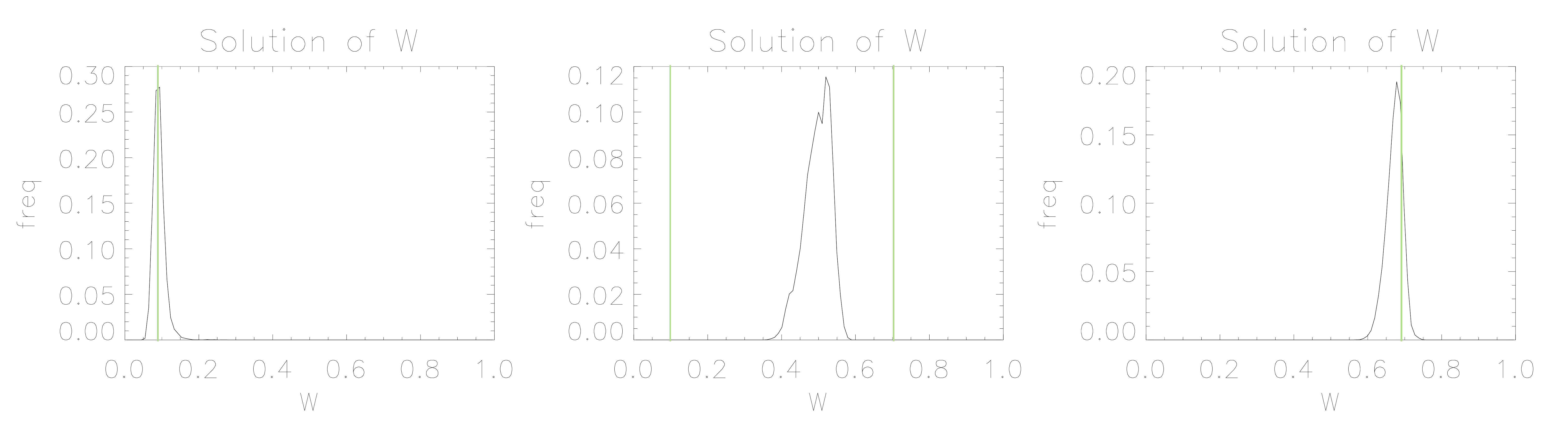}\hfill{}

\caption{Estimated PDF of $\omega$ from 50 geometries of brdf3 (left) and
brdf4 (right) and from the mixture (center), corresponding to figure
\ref{fig:brdf135-brdf246_Refnorm-RefnormNOISE_VS_Phase} middle right
(noisy case). Green lines locate the true values. The concatenation
of two photometric situations (low single scattering albedo $\omega=$0.1
for brdf1 and high single scattering albedo $\omega=$0.7 brdf4) leads
to a misinterpretation of an intermediate photometric situation. \label{fig:brdf3-brdf4_w}}
\end{figure}

\begin{figure}
\hfill{}\includegraphics[width=1\textwidth]{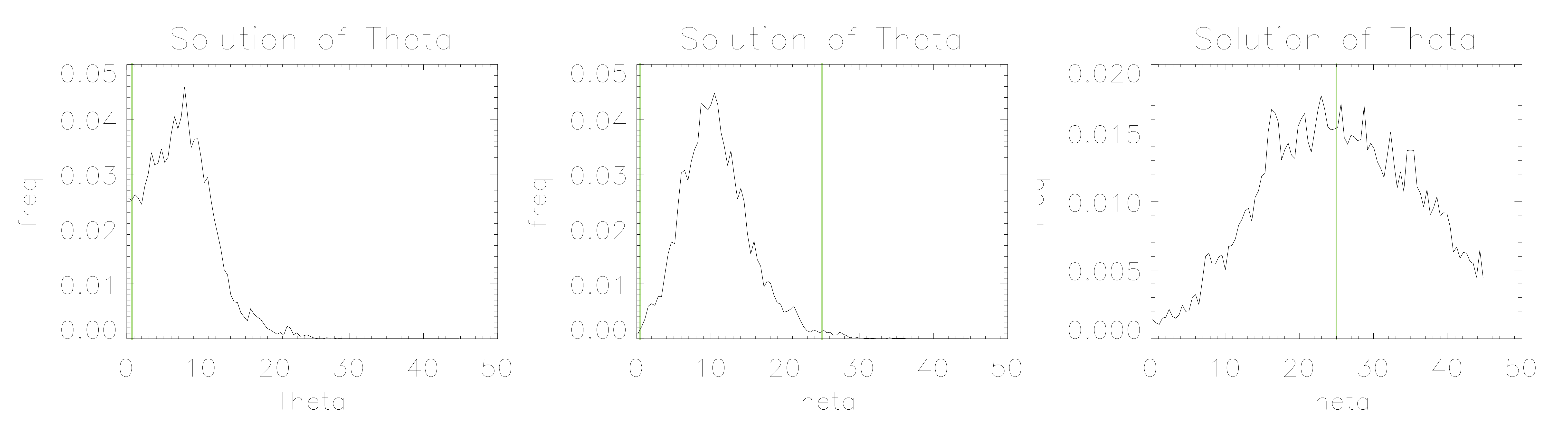}\hfill{}

\caption{Estimated PDF of from 50 geometries of brdf5 (left) and brdf6 (right)
and from the mixture (center), corresponding to figure \ref{fig:brdf135-brdf246_Refnorm-RefnormNOISE_VS_Phase}
bottom right (noisy case). Green lines locate the true values. The
concatenation of two photometric situations (low roughness $\bar{\theta}=$0.5
brdf5 and high roughness $\bar{\theta}=$25 brdf6) leads to a misinterpretation.
\label{fig:brdf5-brdf6_theta}}
\end{figure}

The results of the inversion on the two different phase functions
are shown in Figure \ref{fig:brdf1-brdf2_b_VS_c}. The pure cases
of brdf1 and brdf2 are able to constrain the true values, with higher
uncertainties on the brdf1 case. Broad backscattering is more difficult
to characterize than narrow forward behavior, at this noise level.
Estimation from the mixture of the two data sets yields ``intermediate''
parameter values that do not correspond to any of the two true ones.
Figures \ref{fig:brdf3-brdf4_w} and \ref{fig:brdf5-brdf6_theta}
similarly show the estimated PDFs for the single scattering albedo
and the roughness parameter, respectively: distribution modes are
close to the true values in the two single-photometry cases, but both
lead to erroneous intermediate results with the mixed data set.

Interpretation of the results by analyzing the shape of estimated
PDF may therefore lead to erroneous conclusions, because the model
(assuming that all data come from one single parameter set) is unadapted
to the data. All these figures clearly show that the naive interpretation
of the estimated PDF is not relevant for our problem.

\subsubsection*{Proposed $\chi^{2}$ approach}

We now apply the separability strategy described in section \ref{subsec:Strategy-for-separability-chi2}.
We select the sample with highest likelihood $L$ among the $10^{5}$
trials generated by the MCMC algorithm. 

Table \ref{tab:brf1-brf2_chii} presents the $\chi^{2}$ values obtained
for the three previously described cases, in the presence of noise.
The best fit is noted with tilde (e.g., $\tilde{\chi}^{2}$). It shows
that the homogeneous case of 50 directions is always acceptable: $\tilde{\chi}^{2}\leq\chi_{lim}^{2}$
for all sets of homogeneous data. On the contrary, heterogeneous data
with different$b/c$ and $\omega$ are rejected. Let us remark that
discriminating between homogeneity and heterogeneity in $b/c$ heterogeneous
data was not possible from a simple visual inspection of figure \ref{fig:brdf135-brdf246_Refnorm-RefnormNOISE_VS_Phase}
(top right). However, for $\bar{\theta}$, our $\chi^{2}$ analysis
does not make it possible to decipher a noise effect from a superposition
effect.

Table \ref{tab:brf1-brf2_chii_50_random-noise} shows similar results,
averaged over 50 random noise occurrences, with the same geometry
as defined in Figure \ref{fig:Polarplot-100random} and Table \ref{tab:brf1-brf2_chii}.
Our $\chi^{2}$ analysis allows us to reject homogeneity of $b/c$
and $\omega$ data in all heterogeneous tested cases. For $\bar{\theta}$,
some instances give more favorable results than in the previous example
(30\% of correct decisions), but heterogeneity still remains mostly
undetectable. Only 2\% ($b/c)$ and 10\% ($\omega$ and $\bar{\theta}$)
of homogeneous cases are erroneously rejected. Note that one would
expect a 5\% rejection rate. However, this is an asymptotic rate,
whereas only 50 tests were performed here. Therefore, measured rates
between 2\% to 10\% are still consistent with the theoretical 5\%
value.

Table \ref{tab:brf1-brf2_chii_50_random-geom} finally presents results
averaged over 50 different random geometries, sampled uniformly in
the upper hemisphere. The same conclusions can be drawn regarding
the geometries: 100 random directions are enough to decipher $b/c$
and $\omega$ (100\% of right conclusion), but not $\bar{\theta}$
(30\% of right conclusion). This last test show that all results presented
in this section are thus representative of 100 random directions set.

\begin{table}
\begin{centering}
\begin{tabular}{|c|c|c|c|c|}
\hline 
 & \multicolumn{1}{c|}{$b$/$c$ } & \multicolumn{1}{c|}{$\omega$ } & \multicolumn{1}{c|}{$\bar{\theta}$ } & $\chi_{lim}^{2}$\tabularnewline
\hline 
100superpos & \textbf{164} & \textbf{3606} & 95.3 & 118 (Nb = 100 - 6)\tabularnewline
\hline 
50brdf1,3,5 & 44.0 & 44.8 & 44.7 & 60.5 (Nb = 50 - 6)\tabularnewline
\hline 
50brdf2,4,6 & 37.2 & 38.7 & 36.9 & 60.5 (Nb = 50 - 6)\tabularnewline
\hline 
\end{tabular}
\par\end{centering}
\caption{$\tilde{\chi}^{2}$ value for the best solution over the $10^{5}$
samples generated by the MCMC algorithm, for experiments plotted in
Figure \ref{fig:brdf135-brdf246_Refnorm-RefnormNOISE_VS_Phase}. In
the last column, the maximum acceptable $\chi_{lim}^{2}$ is indicated
with a confidence level of 95\%. Nb indicates the number of degrees
of freedom in the $\chi^{2}$ distribution. \label{tab:brf1-brf2_chii}}
\end{table}

\begin{table}
\begin{centering}
\begin{tabular}{|c||c||c|c||c|c||c|}
\hline 
 & \multicolumn{2}{c|}{$b$/$c$ } & \multicolumn{2}{c|}{$\omega$ } & \multicolumn{2}{c|}{$\bar{\theta}$ }\tabularnewline
\hline 
 & $\tilde{\chi}^{2}$  & $R(\tilde{\chi}^{2}>\chi_{lim}^{2})$  & $\tilde{\chi}^{2}$  & $R(\tilde{\chi}^{2}>\chi_{lim}^{2})$  & $\tilde{\chi}^{2}$  & $R(\tilde{\chi}^{2}>\chi_{lim}^{2})$ \tabularnewline
\hline 
\hline 
100superpos & $170_{119}^{214}\pm20$ & \textbf{100\%} & $3621_{3489}^{3745}\pm48$ & \textbf{100\%} & $111_{71}^{149}\pm19$ & 30\%\tabularnewline
\hline 
50brdf1,3,5 & $44_{28}^{68}\pm9$ & 2\% & $45_{29}^{69}\pm9$ & 10\% & $46_{22}^{79}\pm11$ & 10\%\tabularnewline
\hline 
50brdf2,4,6 & $45_{30}^{64}\pm9$ & 2\% & $49_{30}^{92}\pm12$ & 10\% & $47_{23}^{80}\pm12$ & 10\%\tabularnewline
\hline 
\end{tabular}
\par\end{centering}
\caption{$\tilde{\chi}^{2}$ values of similar problems to the one in Table
\ref{tab:brf1-brf2_chii}, averaged over 50 random noise occurrences
(same geometry as table \ref{tab:brf1-brf2_chii}). For each case,
the left column gives the average and standard deviation of best fit
$\tilde{\chi}^{2}$ is provided, together with their minimum (index)
and maximum (exponent) values. The right column gives the ratio $R$
of instances exceeding the maximum acceptable $\chi_{lim}^{2}$. \label{tab:brf1-brf2_chii_50_random-noise}}
\end{table}

\begin{table}
\begin{centering}
\begin{tabular}{|c||c|c||c|c||c|c|}
\hline 
 & \multicolumn{2}{c||}{$b$/$c$ } & \multicolumn{2}{c||}{$\omega$ } & \multicolumn{2}{c|}{$\bar{\theta}$ }\tabularnewline
\hline 
 & $\tilde{\chi}^{2}$  & $R(\tilde{\chi}^{2}>\chi_{lim}^{2})$  & $\tilde{\chi}^{2}$  & $R(\tilde{\chi}^{2}>\chi_{lim}^{2})$  & $\tilde{\chi}^{2}$  & $R(\tilde{\chi}^{2}>\chi_{lim}^{2})$ \tabularnewline
\hline 
\hline 
100superpos & $183_{122}^{248}\pm27$ & \textbf{100\%} & $3522_{3330}^{3630}\pm75$ & \textbf{100\%} & $109_{78}^{155}\pm20$ & 34\%\tabularnewline
\hline 
50brdf1,3,5 & $46_{26}^{77}\pm11$ & 10\% & $45_{29}^{69}\pm9$ & 10\% & $47_{31}^{75}\pm11$ & 12\%\tabularnewline
\hline 
50brdf2,4,6 & $48_{28}^{68}\pm9$ & 10\% & $49_{30}^{92}\pm12$ & 10\% & $47_{27}^{80}\pm11$ & 12\%\tabularnewline
\hline 
\end{tabular}
\par\end{centering}
\caption{Similar results as in Table \ref{tab:brf1-brf2_chii_50_random-noise},
but results are averaged over 50 random geometries \label{tab:brf1-brf2_chii_50_random-geom}}
\end{table}

\subsubsection{Separability on full Hapke parameter set}

In order to generalize the results described in the previous section,
we propose a test which could be used in more realistic observational
conditions. We consider observations at 10\% noise level, for 100
geometries, randomly taken in the upper half hemisphere, but we do
not know if the dataset is homogeneous or not. In all tests here,
the data set is heterogeneous, with only one Hapke parameter varying
at a time. The full domain of the Hapke parameter set is covered.
We removed the part with $b>0.5$ and $c>0.5$ because no natural
surface seems to be outside the ``hockey cross'' \citep{Hapke_Bidirectionalreflectancespectroscopy_I2012}.

We computed 5,000 test cases for each configuration of the 6 Hapke
parameters,, to which the proposed $\chi^{2}$ strategy was applied
to test homogeneity. Each case has a different random geometry, a
different noise instance and a different random set of Hapke parameters.

\begin{figure}
\includegraphics[width=1\textwidth]{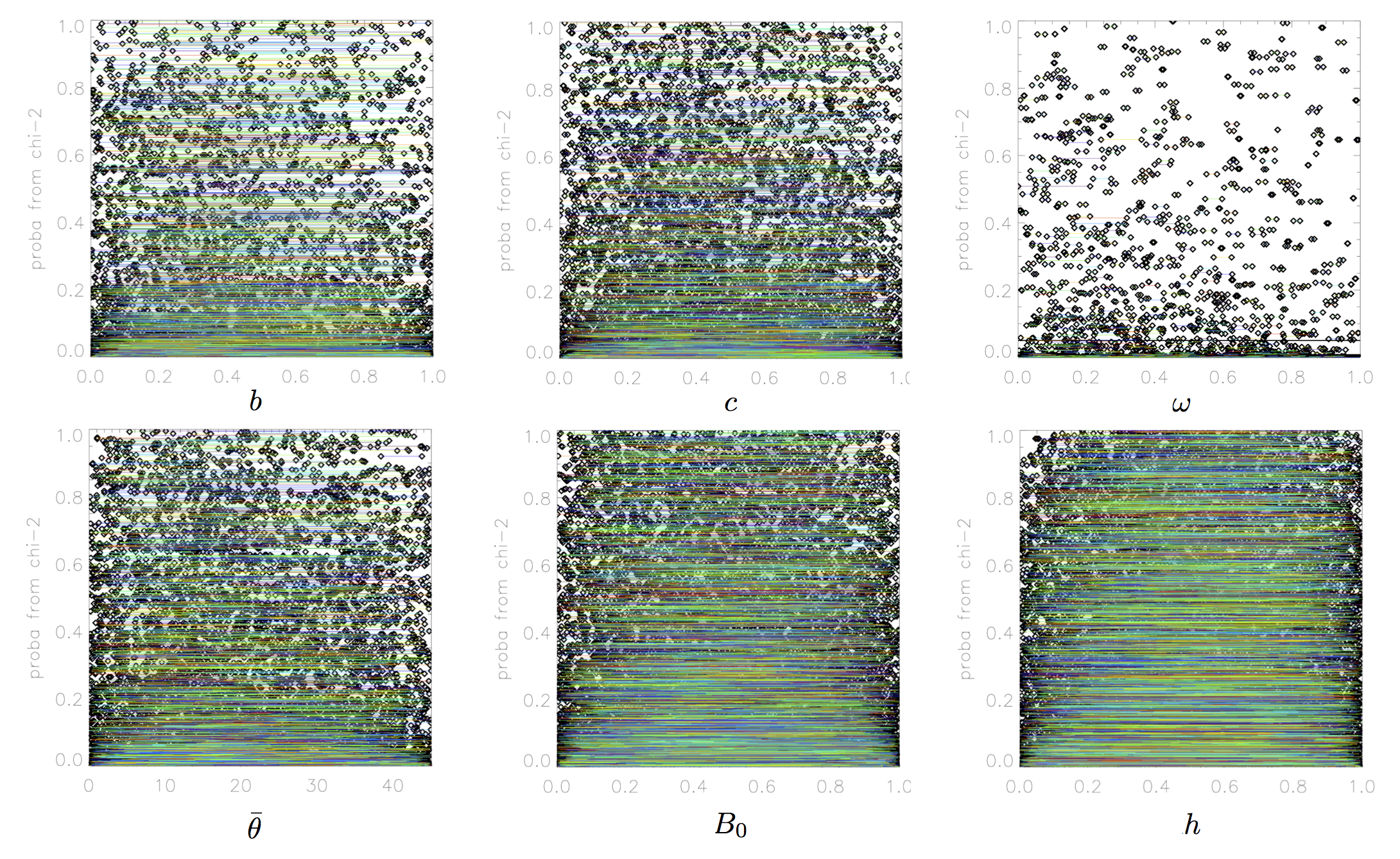}

\caption{Evaluation of our homogeneity analysis. Each graph represents the
probability $\mathscr{P}(\chi^{2}>\tilde{\chi}^{2})$, as a function
of the Hapke parameter. Each couple of test data is plotted by two
diamonds separated by a line. 5,000 examples are plotted for each
graph, with random values of Hapke parameters and random geometries.
The cases that are considered as separable have a probability below
5\%. Thus the points above 5\% will be erroneously classified as homogeneous.
The global rate of misinterpretation is 29\% for $b$, 43\% for $c$,
7\% for $\omega$, 44\% for $\bar{\theta}$, 69\% for $B_{0}$ and
80\% for $h$.}
\label{fig:separability}
\end{figure}

Figure \ref{fig:separability} shows the two values of the parameter
that changed, linked by a straight line, for the 5,000 cases. The
$\tilde{\chi}^{2}$ value was computed and converted into probability
$\mathscr{P}(\chi^{2}>\tilde{\chi}^{2})$, with $100-6=94$ degrees
of freedom. The graph for $\omega$ shows the expected behavior of
separability, since the cases where the probability is higher than
5\% are only cases where the two values of $\omega$ are very close.
For $B_{0}$ and $h$, it seems that both close and far values lead
to high probability and thus the separability is low. $b$, $c$,
and $\bar{\theta}$ are intermediate cases because only few far values
are misinterpreted. The global rate of misinterpretation is 29\% for
$b$, 43\% for $c$, 7\% for $\omega$, 44\% for $\bar{\theta}$,
69\% for $B_{0}$ and 80\% for $h$.

\begin{figure}
\includegraphics[clip,width=1\textwidth]{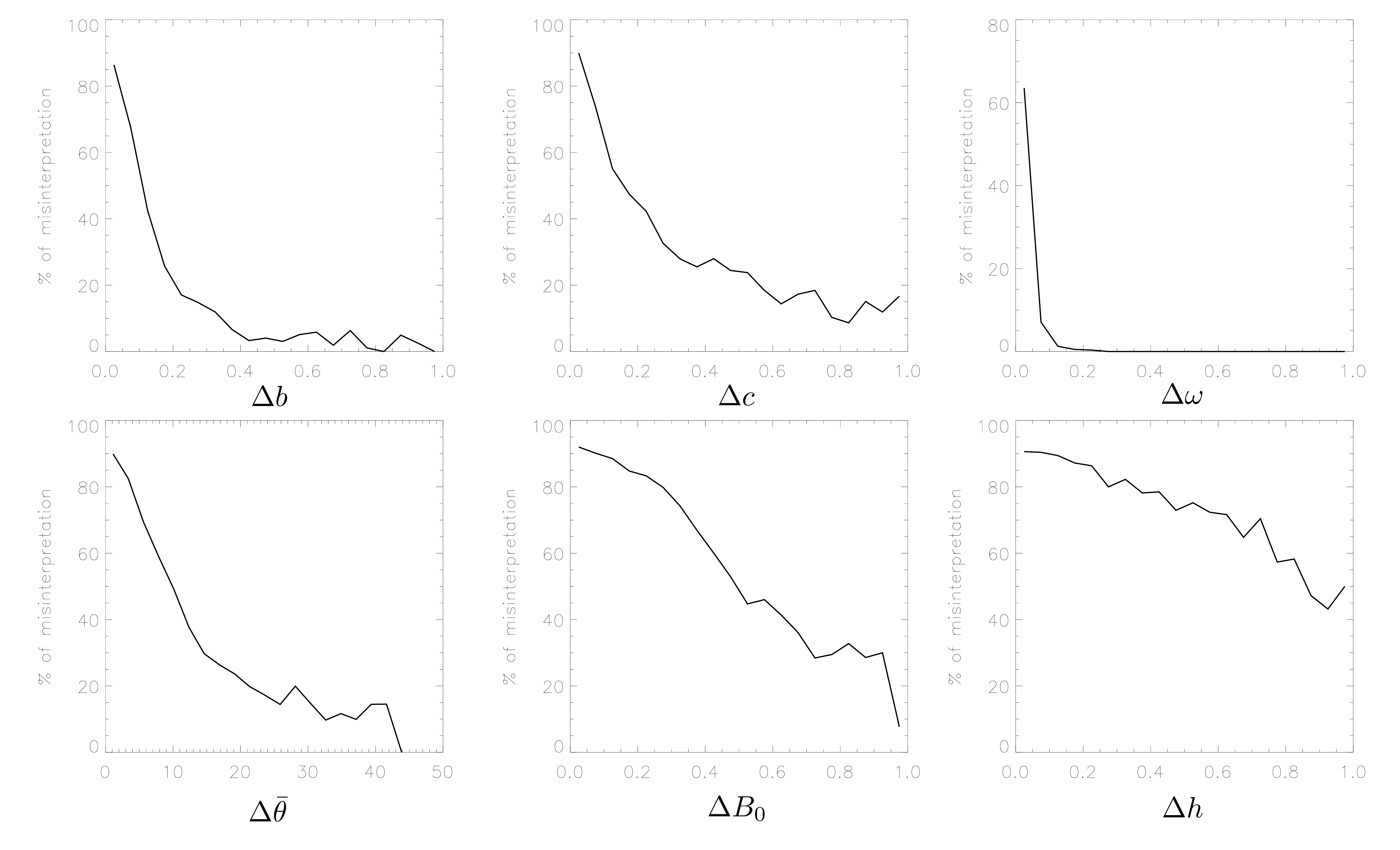}

\caption{Misinterpretation rate of separability as a function of the difference
of the two values of each Hapke parameter. Results are averaged over
5000 random values of Hapke parameters and geometries. The noise level
was set to 10\%. This figure represents the histograms of the data
shown in Fig. \ref{fig:separability}.}
\label{fig:misinterpretation-rate}
\end{figure}

Figure \ref{fig:misinterpretation-rate} shows the misinterpretation
rate for all six parameters, as a function of the difference between
the two values (say, $\Delta m$ for parameter $m$) of the parameter
that changed. Again, the graph for $\omega$ shows the expected behavior
of separability, since the rate of misinterpretation rapidly decreases
as $\Delta\omega$ increases. This rate falls to 0\% for $\Delta\omega\geq0.25$.
This means that whatever the geometry, whatever the photometric parameters,
if $\Delta\omega\geq0.25$, then the proposed strategy always leads
to the conclusion that the dataset is heterogeneous. If we consider
an acceptable misinterpretation rate of 20\%, then the separability
limit is reached for $\Delta b\geq0.2$, $\Delta c\geq0.55$, $\Delta\omega\geq0.05,$
$\Delta\bar{\theta}\geq20.25\text{\textdegree}$, and $\Delta h\geq1$.
This result shows that separability is a very difficult problem, that
has probably been underestimated so far in the literature.

\begin{figure}
\includegraphics[width=1\textwidth]{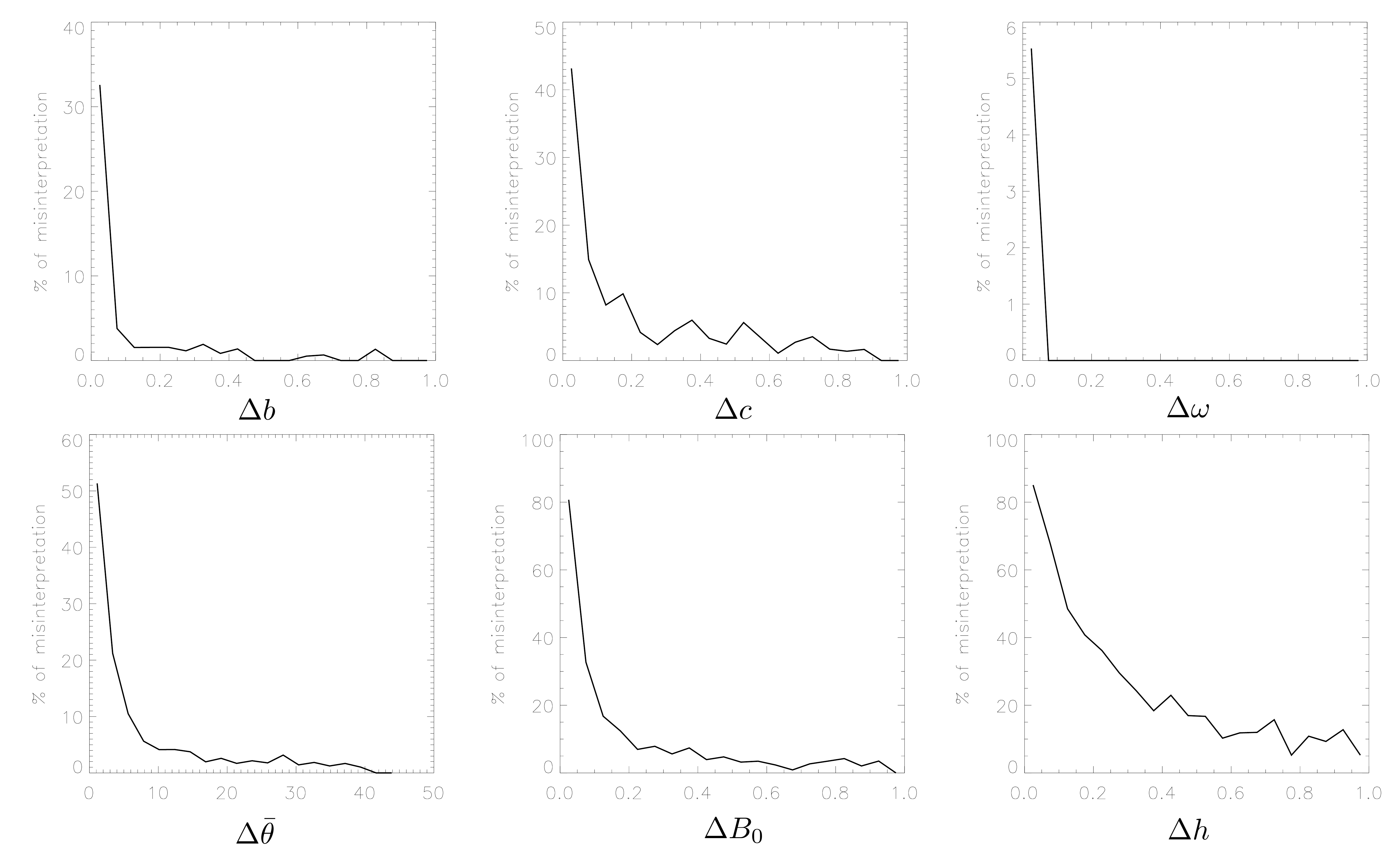}
\caption{Similar results as in Fig. \ref{fig:misinterpretation-rate} but with
the noise level set to 1\% instead of 10\%.}
\label{fig:misinterpretation-rate-1percent}
\end{figure}

We finally note that, if the noise level decreases to 1\%, then all
misinterpretation rates drop down (see figure \ref{fig:misinterpretation-rate-1percent}).
Using the same threshold on $\tilde{\chi}^{2}$ such than $\mathscr{P}(\tilde{\chi}^{2}>\chi_{lim}^{2})=5\%$,
the global rate of misinterpretation is 5\% for $b$, 11\% for $c$,
0.6\% for $\omega$, 10\% for $\bar{\theta}$, 16\% for $B_{0}$ and
35\% for $h$. Those values are significantly lower than those obtained
with 10\% noise, as expected. If we consider a acceptable misinterpretation
rate of 20\%, the separability limit is now reached for $\Delta b\geq0.05$,
$\Delta c\geq0.05$, $\Delta\omega\geq0.05$, $\Delta\bar{\theta}\geq4.5\text{\textdegree}$,
$\Delta B_{0}\geq0.1$ and $\Delta h\geq0.45$.

\section{Discussions and Conclusion}

We proposed a new algorithm to estimate the parameters of the Hapke
model in a Bayesian framework. This method is very fast (a few minutes
for computing the solution on BRDF datasets with tenths of geometries)
and is able to propagate the uncertainties from the measurements to
the parameters, significantly improving the previous implementation
\citep{Schmidt_RealisticuncertaintiesHapke_I2015}. 

We introduced an index to measure the efficiency of a set of geometries
(a collection of directions) to retrieve the proper Hapke parameters.
Using the principal plane with high incidence angle seems the best
solution with a limited number of directions. In particular, such
geometries are better than poorly sampled full BRDF, even with a larger
number of directions, due to the too large phase range. We also noticed
that 5 directions is the minimum number of angular configurations
in the best situation (principal plane) in order to expect well constrained
photometric parameters. We confirmed the power of the Bayesian probabilistic
framework to unravel the effect of mixing different photometric parameters
in a same dataset. For instance, the opposition effect does not seem
to affect the retrieval of $b$/$c$, $\omega$ or $\bar{\theta}$,
at least for the set of tested geometries.

We also proposed a methodology to decipher a difficult observation
condition: aggregation of two photometric surfaces vs. one single
photometric surface. This situation occurs for instance when analyzing
a full planetary body, for which each direction is measured at a different
location. It is also the case when combining in situ observations
from various locations. Our approach is able to answer the following
scientific question: is the collection of data consistent with one
single photometric behavior? In the case of a mixture, the best fit
is often compatible with an intermediate situation that is not present
in the data. In the tested examples, the estimated PDF does not clearly
show two different modes corresponding to the two sets of photometric
parameters. We propose here a simple $\chi^{2}$-based strategy which
provides a confidence test to argue about the homogeneity of the dataset.
From a toy example analysis, it seems that extreme variations of $\omega$
and $b/c$ can be handled with a realistic noise level of 10\%. Heterogeneity
in $\bar{\theta}$ seems very difficult to detect. A more general
analysis demonstrates that this problem can be solved for large differences
for all parameters. The behavior is different for each parameters:
with a 20\% misinterpretation rate, the separability can be achieved
for a 10\% noise level for differences of $\Delta b\geq0.2$, $\Delta c\geq0.55$,
$\Delta\omega\geq0.05,$ $\Delta\bar{\theta}\geq20.25\text{\textdegree}$,
$\Delta B_{0}\geq0.95,$ and $\Delta h\geq1$. With 1\% noise, the
separability can be achieved for a difference of $\Delta b\geq0.05$,
$\Delta c\geq0.05$, $\Delta\omega\geq0.05$, $\Delta\bar{\theta}\geq4.5\text{\textdegree}$,
$\Delta B_{0}\geq0.1$ and $\Delta h\geq0.45$.

The tools proposed and validated here on synthetic tests should now
be applied on real data originating from various planetary science
cases.

All the aforementioned results are valid for surfaces that are compatible
with the Hapke behavior. If the Hapke model is not able to describe
the surface photometry of a specific medium, or if one is interested
by other BRDF models, then the proposed strategy can still be adapted.
In particular, the MCMC algorithm only relies on the computation of
modeled reflectances for any set of input parameters, whatever the
model complexity. 

Future laboratory data and observation data analysis should be done
using the conclusion of this study, especially considering the definition
of the best geometry. The wavelength dependance of all parameters
should also be addressed, for instance by considering each wavelength
independently \citep{Pilorget_Wavelengthdependencescattering_I2016}.
Since the computation time has been significantly reduced, this problem
could be solved easily by running this algorithm for each wavelength
independently. More complex analyses, using an improved Hapke model
including modeled wavelength dependance of parameters, should also
be developed in the future. Last, when a set of directions is clearly
not compatible with one single photometric behavior, new methods should
be developed. In particular, considering a model mixture which would
split the data into different classes (with a priori an unknown number
of classes), combined with adapted estimation and classification algorithms
(e.g. based on reversible jump MCMC \citep{Green_ReversiblejumpMarkov_B1995}),
is an ambitious perspective to these works.

\subsubsection*{Acknowledgements}

We acknowledge support from the ``Institut National des Sciences
de l'Univers'' (INSU), the \textquotedbl{}Centre National de la Recherche
Scientifique\textquotedbl{} (CNRS) and \textquotedbl{}Centre National
d'Etudes Spatiales\textquotedbl{} (CNES) through the \textquotedbl{}Programme
National de Plan\'etologie\textquotedbl{}, the MEX/OMEGA and the
MEX/PFS programs. The project Multiplaneto has been granted in
2015 and 2016 by D\'efi Imag\textquoteright In/CNRS. 

\appendix

\section{Monte Carlo Markov Chain algorithm\label{sec:Appendix-Monte-Carlo-Markov}}

In the previous implementation, the generation of new candidates was
performed using a uniform distribution in the parameter space (see
pseudo-code \ref{alg:Previous-MCMC-sampling-1}). Consequently, many
proposals were drawn in regions with very low probability density
values, that were rejected. As a result, a very high number of random
drawings was necessary in order to sample the PDF of interest. In
the new version, we propose to generate the candidate using three
possible options (see pseudo-code \ref{alg:New-MCMC-sampling-1}):
1: with a uniform distribution as previously in order to be sure to
cover the full parameter space, but only in 20\% of the cases in average;
2: in the large neighborhood of the previous sample (Gaussian random
walk with large standard deviation: 10\% of the parameter space) in
order to escape from possible local minima regions; 3: in the close
neighboring of the previous sample (Gaussian random walk with small
standard deviation, set to 0.1\% of the parameter space) in order
to refine local exploration.

Another change is that we now keep systematically the sample (either
the candidate or the previous one, using the Hastings-Metropolis rule).
This strategy improves convergence in the case of narrow PDFs.

Let us remark that the previous implementation was slightly erroneous:
a sample must be kept at each iteration (either $m_{i+1}=m^{\star}$
or $m_{i+1}=m_{i})$ in order to achieve asymptotic distribution of
the samples according to the PDF $\sigma_{M}$. In practice, the results
are not significantly affected by this error. We finally note that
the expression of the acceptance probability $\rho$ in the algorithm
reduces to the ratio $\sigma_{M}(m^{\star})/\sigma_{M}(m_{i})$ since
all distributions (the distribution used for drawing $m^{\star}$
as a function of $m_{i}$) are symmetric. Therefore, the ratio that
may appear in the expression of $\rho$ (see for example \citep{Robert_MonteCarlo_book2005})
equals 1.
\begin{algorithm}
Initialization ($m_{0}$, $L_{0}$,$i=0$)

while $i\leq1000$

\quad{}\quad{}i) random generation of a candidate $m^{\star}$ sampled
uniformly in $M$

\quad{}\quad{}ii) computation of the interest probability density
function $\sigma_{M}(m^{\star})$\quad{}\quad{}iii) accept $m_{i+1}=m^{\star}$
using the Hastings-Metropolis rule, that is, accept it with probability
$\rho=\frac{\sigma_{M}(m^{\star})}{\sigma_{M}(m_{i})}$. 

\quad{}\quad{}If the candidate is accepted, then ; otherwise go
back to i)

\caption{Previous MCMC sampling\label{alg:Previous-MCMC-sampling-1}}
\end{algorithm}

\begin{algorithm}
Initialization ($m_{0}$, $L_{0}$)

for $i=1:10^{5}$

\quad{}\quad{}random generation of a candidate $m^{\star}$:

\quad{}\quad{}\quad{}\quad{}for each parameter from $(\omega,b,c,\bar{\theta},h,B_{0})$
independently

\quad{}\quad{}\quad{}\quad{}\quad{}1 with probability $\frac{1}{5}$,
draw $m^{\star}$ uniformly in $M$ 

\quad{}\quad{}\quad{}\quad{}\quad{}2 with probability $\frac{2}{5}$,
draw $m^{\star}$ = $m_{i}$ + large random step

\quad{}\quad{}\quad{}\quad{}\quad{}3 with probability $\frac{2}{5}$,
draw $m^{\star}$ = $m_{i}$ + small random step

\quad{}\quad{}computation of the interest probability density function
$\sigma_{M}(m^{\star})$

\quad{}\quad{}\quad{}accept the candidate $m_{i+1}=m^{\star}$
using the Hastings-Metropolis rule, that is, accept it with probability
$\rho=\frac{\sigma_{M}(m^{\star})}{\sigma_{M}(m_{i})}$

\quad{}\quad{}\quad{} otherwise keep $m_{i+1}=m_{i}$ 

\caption{New MCMC sampling\label{alg:New-MCMC-sampling-1}}
\end{algorithm}

\subsubsection*{Reference}

\bibliographystyle{elsarticle-harv}

\end{document}